# Computational multi-spectral video imaging

**PENG WANG,**[1,&] **RAJESH MENON**[1,*]

[1]*Department of Electrical and Computer Engineering, University of Utah, Salt Lake City, UT 84112*
[&]*Current address: Department of Medical Engineering, California Institute of Technology, Pasadena, CA 91125*
*Corresponding author: rmenon@eng.utah.edu*

**Multi-spectral imagers reveal information unperceivable to humans and conventional cameras. Here, we demonstrate a compact single-shot multi-spectral video-imaging camera by placing a micro-structured diffractive filter in close proximity to the image sensor. The diffractive filter converts spectral information to a spatial code on the sensor pixels. Following a calibration step, this code can be inverted via regularization-based linear algebra, to compute the multi-spectral image. We experimentally demonstrated spectral resolution of 9.6nm within the visible band (430nm to 718nm). We further show that the spatial resolution is enhanced by over 30% compared to the case without the diffractive filter. We also demonstrate Vis-IR imaging with the same sensor. Since no absorptive color filters are utilized, sensitivity is preserved as well. Finally, the diffractive filters can be easily manufactured using optical lithography and replication techniques.**

## 1. INTRODUCTION

Traditional imaging systems map one point in the object space to one point in the image space [1]. The spatial extension of the imaged point, the point-spread function (PSF), is essentially determined by far-field diffraction and aberrations present in the system. Historically, advanced lens design and manufacturing techniques were developed to minimize all kinds of aberrations to achieve the diffraction-limited PSF [1]. Over the past decades, several methods have been extensively explored to resolve sub-diffraction features in super-resolution microscopy [2], by either shrinking the physical dimension of the PSF [3,4] or by using statistical estimation with pre-knowledge on the shape of the PSF [5,6]. However, they are not applicable to traditional photography systems.

Furthermore, electronic sensors can only detect light intensity. In order to distinguish colors, an absorbing color-filter array (generally called the Bayer filter) is placed on top of the sensor [7]. Typically, only three colors (blue, green and red) are measured. However, natural scenes contain multi-spectral information, which can be valuable for numerous machine-vision applications. Conventional multi-spectral imagers (MSI) are expensive and cumbersome. A common MSI operates in a push-broom manner and utilizes a prism or grating to disperse light [8]. Although the optical design can be simple, its applications are limited to scenarios where the MSI is scanned relative to the object such as on a satellite or on a conveyor belt. A second category of MSI employs either liquid-crystal tunable filters or acousto-optic tunable filters to modulate the input spectrum over time [9,10]. All these techniques scan multiple 2D projections (($x, \lambda$) or ($x, y$)) to acquire 3D multi-spectral data ($x, y, \lambda$), and here, are slow. There is a need to acquire the 3D multi-spectral data in only one shot [11,12], particularly to avoid motion artifacts.

Single-shot multi-spectral imagers based on coded apertures have demonstrated reasonable image quality and spectral resolution [11,13]. However, the introduction of a patterned absorbing aperture, a dispersive element (prism) and relay optics increases the system size and complexity. Its spatial and spectral resolutions are dependent upon the sparsity of objects, due to the adoption of the concept of compressed sensing. Recently, multi-spectral sensors based on a tiled bandpass-filter array have become commercially available. Fabry-Perot (FP) resonators are integrated on CMOS sensors to achieve spectral selectivity [14-16]. These not only require expensive fabrication steps, but also exhibit very poor sensitivity. These resonator-based filters may be replaced by plasmonics-based alternatives. But such alternatives incorporate sub-wavelength structures that are difficult to manufacture, [17,18] and these also suffer from low sensitivity due to parasitic absorption losses. Most importantly, tiled-filter-based imagers trade-off spatial resolution with spectral resolution.

In this article, we convert a conventional camera into a single-shot multi-spectral video imager by inserting a thin diffractive filter in the near vicinity of the image sensor. By applying appropriate computational algorithms, we are able to achieve multi-spectral video imaging with improved spatial resolution. The imaging speed is limited by the frame rate of the camera. Compared to our previous work [19], the current implementation is generalized to use any imaging lens to increase the field of view, is able to achieve high resolution in both spectrum and space, is able to image in both the visible and NIR regimes, perform computational refocusing and computationally trade-off spectral resolution against the field of view without changing the hardware. It is important to point out that there is no violation of the conservation of entropy because the total information acquired (the size of the raw sensor frame) constrains the achievable spectral resolution and the field of view. In our proof-of-concept experiments, we acquire a raw sensor frame of size 150×150 pixels, from which we can reconstruct multi-spectral frames of either 30×30 pixels×25 bands or 50×50 pixels×9 bands. Finally, we also describe experimental and numerical analyses of the noise performance, modulation transfer function and depth of focus of the imaging system.

Our imager is schematically described in Fig. 1(a). In a conventional imaging system under geometrical optics, the lens images single points (A & B) in the object plane (XY) onto single points (A' & B') in the image plane (X'Y'). In our system, we insert a thin diffractive filter (DF) in front of the image plane. Therefore, before converging to points A' and B', light is diffracted by the diffractive filter. The diffraction patterns received by the sensor, represented by the blue and red circles in Fig. 1(a), are wavelength-dependent. The diffraction patterns also vary according to different spatial locations of their original points in the object plane (A & B). The centers of the diffraction patterns, blue and red circles, still coincide with the points A' and B'. Therefore, one wavelength at one object point uniquely corresponds to one diffraction pattern in the sensor plane. Diffractive optics is known to suffer from chromatic aberration and various methods were developed for broadband operation by nonlinear optimization [20], metasurfaces [21] and joint design between diffractive optic and image processing [22]. In contrast, here we exploit this chromatic aberration to enable multispectral imaging. Alternatively, this also can be considered as a way to correct chromatic aberration using only computation.

## 2. WORKING PRINCIPLE

According to the description above, the intensity distribution of the recorded image is basically a linear combination of the diffraction patterns of all the wavelengths at all the spatial points that contribute to the image. This can be expressed as a matrix multiplication: $\mathbf{I}=\mathbf{AS}$, where $\mathbf{S}(x, y, \lambda)$ is the unknown 3D multi-spectral data cube, $\mathbf{I}(x', y')$ is the intensity distribution of the 2D image on the sensor and $\mathbf{A}(x', y'; x, y, \lambda)$ is the 5D matrix representing the response due to the diffractive filter. In our preliminary implementation, the object plane is discretized to 30 by 30 points, spaced by $\Delta X=\Delta Y=120\mu m$ (purple grid in Fig. 1(a)). At each point, we compute 25 wavelengths between 430nm and 718nm in steps of 12nm. To experimentally calibrate the matrix $\mathbf{A}$, we mounted a pinhole on a 2D motorized stage to scan across the object grid and illuminated the pinhole by a supercontinuum source equipped with a tunable bandpass filter that selects the wavelengths. Retrieving $\mathbf{S}$ from $\mathbf{A}$ and $\mathbf{I}$ is a typical inverse problem ($\mathbf{S}=\mathbf{A}^{-1}\mathbf{I}$) that can be readily solved via Tikhonov regularization [23]. Iterative algorithms are slow to converge and thus are not used here [24,25]. Note that each frame is sufficient to obtain the multi-spectral data for that frame and by piecing together multiple frames, multi-spectral video can be readily generated. We note that the regularization parameter is independent of the scene details and is the same for all frames. Video imaging is identical to single-frame imaging as long as the computation can be performed at the video rate. The idea of extracting high-dimensional information from multiplexed data is adopted from previous works in computational spectroscopy [26–28].

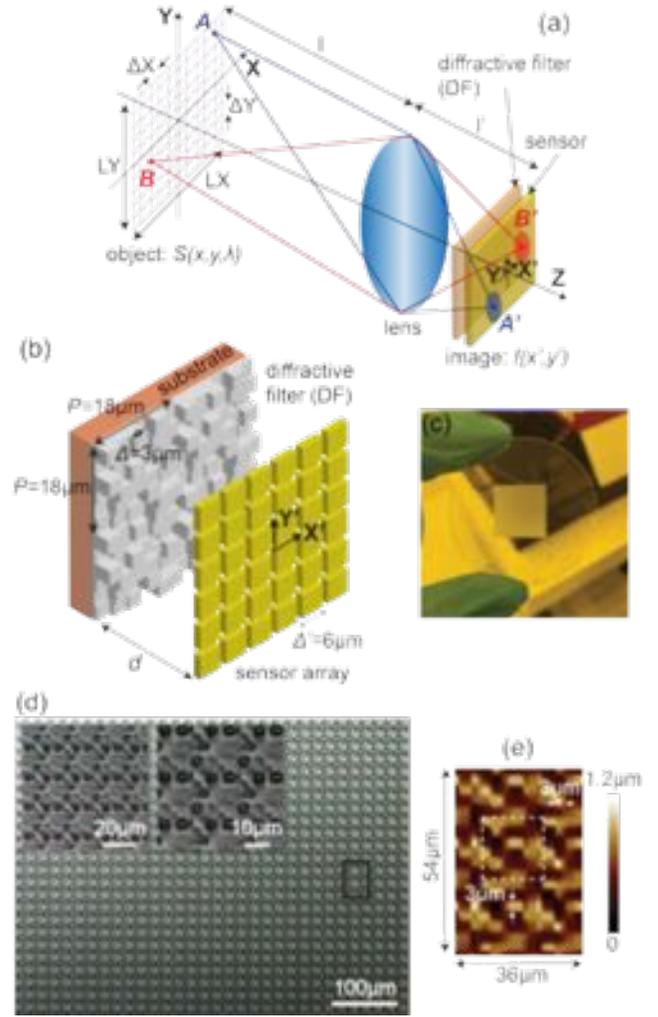

**Fig. 1.** (a) Schematic of the single-shot multi-spectral imager. A diffractive filter is placed in close proximity to the sensor. Due to diffraction through the filter, object points A and B are imaged to diffraction patterns (blue and red circles) surrounding points A' and B' on the sensor. The diffraction patterns depend on the wavelength and the spatial location of the object point. (b) Schematic of the assembly comprised of the diffractive-filter (DF) and the sensor array. The DF has features of width 3μm and period 18μm. The monochrome CMOS sensor has pixel size of 6μm. (c) A photograph of the fabricated DF on glass substrate. (d) Micrographs of the fabricated DF. Oblique illumination is applied to enhance contrast (insets: images with larger magnifications). (e) An atomic-force measurement of the DF delimited by the black box in (d). The structure has maximum height of 1.2μm. Pixel size of 3μm is labeled and one period is enclosed by the white dashed box.

The schematic of the diffractive filter (DF) is depicted in Fig. 1(b). It is a 2D multi-level structure comprised of a super-lattice of period $P$=18μm and constituent square pixel of size, $\Delta$=3μm. Each pixel is quantized into various height levels and the maximum height is set to 1.2μm for ease of fabrication. Note that periodicity is not necessary in this application. For this preliminary demonstration, we simply chose the pixel heights from a pseudo-random set. The gap, $d$ between the DF and the sensor is approximately 0.5mm. We used a monochrome CMOS sensor (DMM22BUC03-ML, The Imaging Source) with sensor-pixel size = 6μm. The DF is patterned in a transparent dielectric material (Shipley 1813 photoresist) coated on fused silica substrate via gray-scale lithography [29-31] (see Fig 1(c)). Figure 1(d) provides optical micrographs of the DF at three magnifications (VHX-5000, Keyence). The shadows in the images are created by oblique illumination to emphasize the three-dimensional profile. An atomic force microscopy (AFM) image is shown in Fig. 1(e), where the white dashed square encloses one period and the 3μm features are clearly marked. A commercial lens (MVL12WA, Thorlabs, resolution of ~600 cycles/mm) is placed in front of the DF.

Our first step is to measure the matrix **A**. This is achieved via a calibration setup, the details of which are included in the Supplementary Information [32]. A super-continuum source (EXW6, NKT Photonics) is collimated and expanded, then illuminates the pinhole (diameter, $\varphi$=150μm). In order to ensure that the illumination overfills the aperture of the lens, the pinhole is mounted at the focus of an achromatic lens and a diffuser is glued to its back. The lens magnification is ~24 and the image of the pinhole is smaller than the diffraction-limited spot-size. The pinhole is stepped along the object grid (Fig. 1(a)) using a 2D stage (Z825B, Thorlabs). A tunable bandpass filter (VARIA, NKT Photonics) is used to select the wavelength for illumination. We utilized a bandwidth of 12nm for our experiments. Exemplary measured values of $\mathbf{A}(x', y'; x, y, \lambda)$ are plotted in Fig. 2(a). They are at five different object point locations $(x, y)$ and four different wavelengths $(\lambda)$. Note that this calibration only needs to be carried out once and the data can be used for all frame reconstructions. As mentioned earlier, the data cube **S** has a dimension of 30×30×25=22500. In order to solve the inverse easily, **A** is defined as a square matrix and thus the raw image **I** needs to have a dimension of 150×150 pixels (see Supplementary Information [32]). They are cropped from the original camera images. From Fig. 2(a), it is clear that the diffraction patterns change with both wavelength and location. In other words, the point-spread function is both spatially and spectrally variant.

Solving the inverse problem using regularization starts with singular-value-decomposition (SVD) of matrix $\mathbf{A}=\mathbf{U\Sigma V}$. $\mathbf{\Sigma}$ is a diagonal matrix with the singular values as its diagonal elements, arranged in a descending manner $(\sigma_1 \geq \sigma_2 \geq \sigma_3 \geq ... \geq \sigma_k)$. The columns of **U** and **V** matrices $(u_1, u_2, u_3...u_k,$ and $v_1, v_2, v_3...v_k)$ contain the $k$×1 left and right singular vectors, respectively. From the Fourier expansion point-of-view, forward problem diminishes the high-frequency components in $u_i$ and $v_k$, while the inverse process attempts to magnify those high-frequency parts. The regularization stabilizes the problem by minimizing both the residual norm $||\mathbf{AS}\text{-}\mathbf{I}||_2$ and the solution norm $||\mathbf{S}||_2$. Usually, they are balanced by a regularization parameter $\omega$.

$$\min\left\{\left\|\mathbf{AS}-\mathbf{I}\right\|_2^2 + \omega^2 \left\|\mathbf{S}\right\|_2^2\right\}. \quad (1)$$

Mathematically speaking, it can be equally formulated as applying filter factors to solution vectors [23,32]:

$$\mathbf{I}_\omega = \sum_{i=1}^{n} \phi_i^{[\omega]} \frac{u_i^T \mathbf{S}}{\sigma_i} v_i, \quad (2)$$

in which the filter factor is defined as:

$$\phi_i^{[\omega]} = \frac{\sigma_i^2}{\sigma_i^2 + \omega^2}. \quad (3)$$

As in computational spectroscopy [26-28], we can estimate the spatial and spectral resolutions via the cross-correlation between the diffraction patterns at the object coordinates, $(x, y, \lambda)$. For example, spectral correlation can be calculated by:

$$C(x, y, \Delta\lambda) = \left\langle \int \mathbf{A}(x', y'; x, y, \lambda) \cdot \mathbf{A}(x', y'; x, y, \lambda+\Delta\lambda) dx' dy' \right\rangle \quad (4)$$

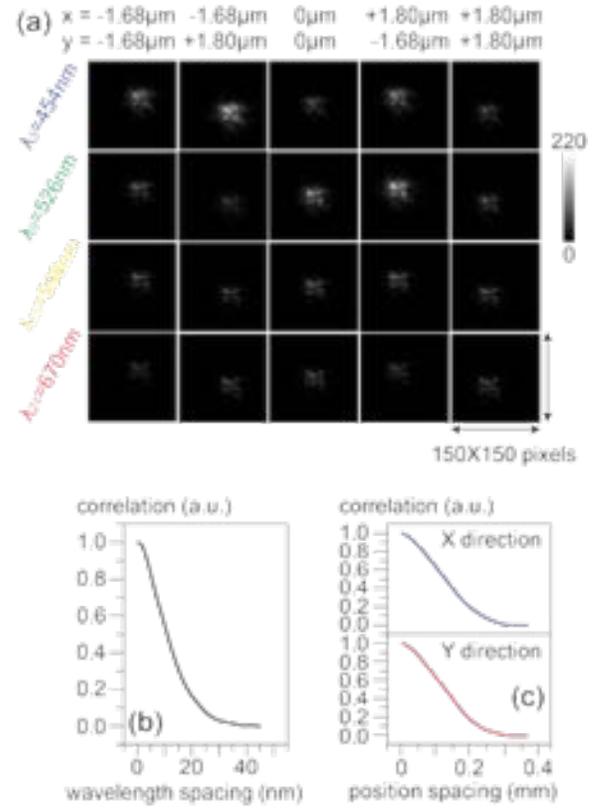

**Fig. 2.** (a) Exemplary measured data for $\mathbf{A}(x', y'; x, y, \lambda)$ at five spatial locations and four wavelengths. Each frame has 150 by 150 pixels. (b) Spectral correlation functions versus wavelength spacing. (c) Spatial correlation functions versus position spacing (top panel: X direction; bottom panel: Y direction).

The bracket <...> represents average over all the wavelengths. Similar expressions can be obtained for spatial correlations. Figures 2(b) and 2(c) plot the spectral and spatial cross-correlation functions of the constructed imaging system versus the sampling in the wavelength and spatial domains. This analysis of resolution is clearly independent of the specific reconstruction algorithm or regularization parameter, and hence, is a fundamental metric of our camera. Here a finer wavelength scan step of 1nm is used to obtain a smooth curve and the calculated spectral resolution, defined as the wavelength spacing when cross-correlation equals 0.5, is ~9.6nm, [26]. Note that this spectral resolution is slightly smaller than the spectral sampling size, 12nm during calibration. The plot in Fig. 2(b) is averaged over all spatial positions. Figure S6 in the Supplementary Information [32] further show that the spectral resolution remains the same over the entire calibration space. Similarly, a finer scan step of 10μm in space is used to compute the spatial cross-correlation, plotted in Fig 2(c). A spatial resolution of ~120μm on the object plane is estimated in both orthogonal directions (X and Y) when cross-correlation equals 0.5. This corresponds to ~5μm resolution on the image plane. Again, Fig. 2(c) is averaged over all wavelengths and a detailed analysis in the Supplementary Information [32] show that the spatial resolution is higher in the blue and green wavelengths than that in the red. These values from Figs. 2(b) and 2(c) are only upper bounds in resolution inherent to the current system, since the final resolutions are also determined by the sampling (step) size during calibration. Intuitively speaking, larger step sizes in the wavelength and spatial domains lead to poorer spectral and spatial resolution, respectively. Since the imaging lens provides a demagnification factor of 24×, the spatial resolution determined by the sensor-pixel size is 6μm on the image plane and is 6μm×24=144μm on the object plane. The simulated spectral and spatial resolutions are somewhat smaller than expected values. This is due to the fact that our image reconstruction is analogous to fitting to a structured PSF, which can be theoretically shown to have a lower Cramer-Rao lower bound and thereby, attain higher resolution. An example of such a technique that achieved higher spatial resolution in localization microscopy has been demonstrated before [33]. Here, we note that the similar principle operates in both the spectral and spatial domains. Note that both resolutions can be tuned by either changing the pinhole diameter in calibration or adjusting the gap between the DF and the sensor [19]. In addition, we can computationally trade-off spectral resolution for field of view, and vice-versa without any change in the hardware (see discussions).

## 3. RESULTS

For all the following imaging experiments, the spatial sampling in the object plane was 120μm during calibration. Various color images were displayed on the LCD screen of an iPhone 6 placed in the object plane. The raw monochrome data, $\mathbf{I}(x', y')$ are captured at such exposure times, which ensure that the sensor pixels are not saturated. A second optical path captured reference RGB images using the same lens and the same sensor chip, but with the conventional Bayer filter (DFM22BUC03-ML, The Imaging Source). Multi-spectral images were reconstructed as described earlier. RGB images were computed from the multi-spectral data for comparison to the reference images, using standard transmission functions for the Bayer-filter array. The results of six different objects are summarized in Fig. 3. The multi-spectral data has 25 wavelengths, ranging from 430nm to 718nm separated by 12nm and arranged into four rows (Figs. 3(a) – (f)), roughly corresponding to blue, green, red and near-infrared from top to bottom. The wavelength for each sub-image is labeled in Fig. 3(d). All the others in Figs. 3(a) – (f) follow the same sequence. The first example is a 2-color letter 'T'. According to the multi-spectral images in Fig. 3(a), the vertical green bar contains signals from λ=514nm to 598nm, while the top red bar contains signals from λ=586nm to 682nm. In the multi-spectral plots of Fig 3(b), the first, the second and the third rows contain blue, green and red spectra of the left, the center and the right parts of the letter 'H', respectively. The plots of near-infrared in the fourth row are blurred and noisy due to low signal of the object (iPhone screen) at those wavelengths. Figure 3(c) is an object of four letters 'U' 'T' 'A' and 'H' of four different colors. The yellow 'H' doesn't come out accurately due to inappropriate white balancing in this demonstration. This is not implemented here in order to show the capability of the imaging system using just reconstruction without post-processing. Any post-processing technique can be readily incorporated for future applications. The letter 'A' in blue channel is not reconstructed properly, compared to the letters of other colors, primarily due to the brightness difference between the 3 RGB color channels in the source (iPhone screen). Figure 3(d) illustrates excellent reconstruction of a single letter 'A' that consists of only one blue color. This indicates that applying high-dynamic range (HDR) algorithms and hardware may help to improve the image quality. These can be adapted for specific applications. The nine-dot-array in Fig. 3(e) has more colors. The center white dot has signals contributed from all the wavelengths except the near-infrared row. The left middle yellow dot contains green and red spectra. The bottom purple dot has blue and red channels but leaves the green channel blank. Blue and green spectra contribute to the cyan dot at the right middle. The reconstructed spectra at the centers of the dots (normalized to its maximum) are plotted in Fig. 4. They match well with the reference spectra measured by a commercial spectrometer (Ocean Optics Jaz) in black lines. The reference spectra are down-sampled to the same spectral resolution with the reconstruction for fair comparison. The average error, calculated as spectrally-integrated error over spectrally-integrated spectrum, is less than 8%. The multi-spectral images of a rainbow are shown in Fig. 3(f). The peak wavelength is red-shifted from left to right as expected.

Our reconstructed images also suggest that we can attain better spatial resolution than the reference color camera. The shapes of the reconstructed color patterns are clearer and narrower than those of the reference. Extra noises in the reconstructed images are ascribed to mechanical drifts and sensor noise in both calibration and raw images. We experimentally confirmed that the average cross-talk between neighboring spectral channels is about -5dB and the spectral cross-talk between channels that are not neighbors is smaller than -10dB (see Supplementary Information [32]). Currently, it takes ~0.08 sec to complete one frame reconstruction using regularization in MATLAB, running on a laptop computer (Intel i7 Core, 16GB RAM). Note that an optimal regularization parameter of $\omega$=3.0 is used in all reconstructions (see Supplementary Information [32] for more details). Too large regularization parameter distorts the signal while too small parameter is unable to sufficiently suppress high-frequency noises. Further optimization of the reconstruction algorithm and hardware implementation will speed this up significantly.

To illustrate multi-spectral video-imaging, we used our camera to capture a test video of a rotating three-color letter 'H', displayed on the iPhone screen as before. The frame rate is 15 fps. The sensor data was processed and reconstructed in real-time. The original video to be displayed (Visualization S1), the raw monochrome video taken by the DF-sensor assembly (Visualization S2), the multi-spectral videos at all wavelengths (Visualization S4) and the synthesized RGB video (Visualization S3) are included as the Supplementary Materials. As expected, the reconstructed RGB video matches the original RGB video on iPhone screen quite well. Note that the computational part of the video imaging is identical to that for still imaging.

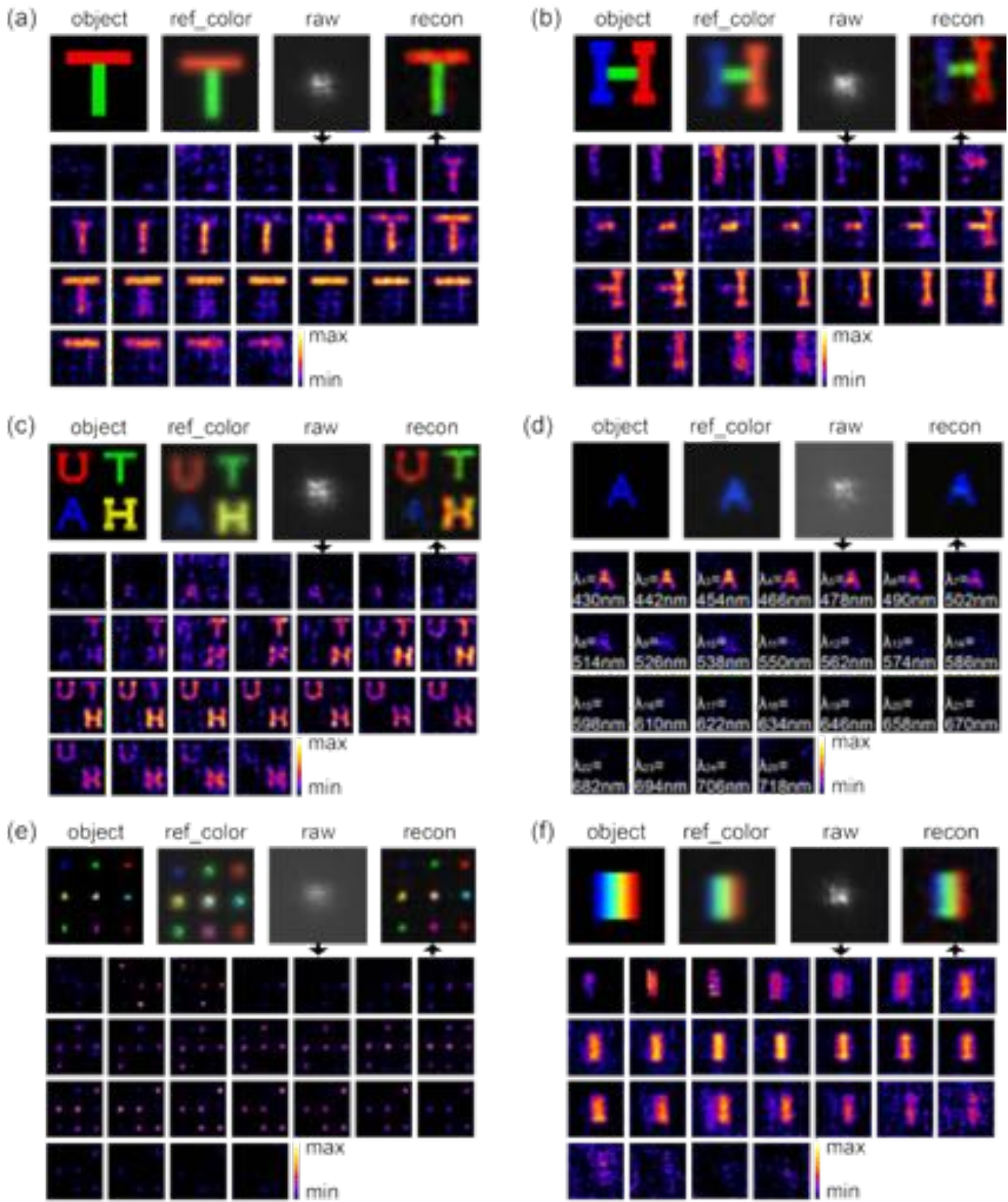

**Fig. 3.** Experimental results of the multi-spectral frames and color images using regularization based on the raw images and the calibrated PSFs. The designed object patterns to be displayed, the reference RGB images (3.6mm by 3.6mm field-of-view, 25 X 25 pixels), the raw monochrome images (150 by 150 sensor pixels) and the reconstructed RGB images (3.6mm by 3.6mm field-of-view) of the test patterns are shown as the top row. The normalized multi-spectral frames (3.6mm by 3.6mm field-of-view) are plotted in the bottom four rows. Six object patterns are tested: (a) 2-color letter 'T'; (b) 3-color letter 'H'; (c) 4-color letters 'UTAH'; (d) 1-color letter 'A'; (e) 7-color dot-array; (f) Rainbow.

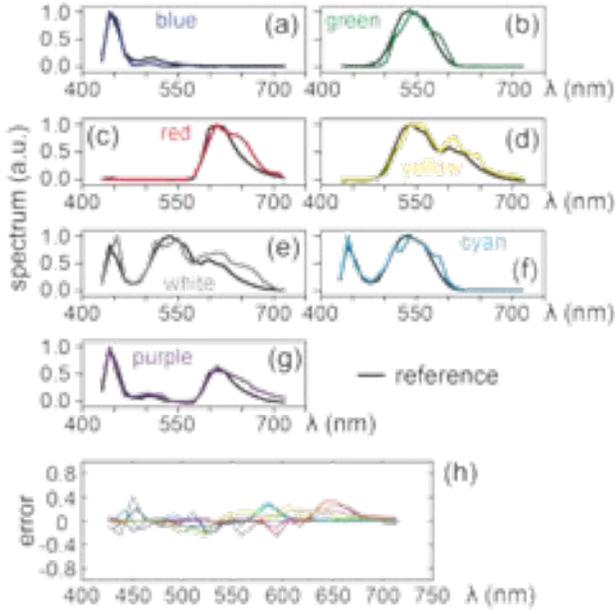

**Fig. 4.** Normalized spectra at the centers of the dots of different colors in Fig. 3(e). The reference spectra measured by a commercial spectrometer are plotted in black solid lines. The spectra are of (a) blue; (b) green; (c) red; (d) yellow; (e) white; (f) cyan and (g) purple colors. (h) Plots of errors between reconstructed and reference spectra. Colors of curves correspond to those of colors in (a)-(g). The errors between reconstructed and reference spectra are less than 8% on average (see Supplementary Information [32]).

## 4. DISCUSSION

To quantify the spatial resolution of our system, we measured the modulation transfer function (MTF) of our camera. The object was comprised of periodic lines of the same color. Since the iPhone 6 screen has resolution of 326ppi (78µm pixel pitch), the minimum period of the test pattern was 2×78µm=156µm (or the maximum spatial frequency, $v$=6.4cycles/mm). The measured MTFs along the X and Y axes, and for the 3 basic colors (blue, green and red) are plotted in Fig. 5(a). From the measured data, we observe that the spatial resolution (defined as the line-spacing where MTF = 0.1) is increased by 43% and 20% along the X and Y axes, respectively, averaged over the 3 colors, when compared to that obtained using the conventional Bayer filter in our reference camera. This is in stark contrast to conventional multi-spectral imagers, where spatial and spectral resolutions are traded off against one another. Figure 5(b) shows some exemplary MTF test patterns and images at five different spatial frequencies and various colors. The reconstructed images have higher contrast when compared to the reference images. Although our default data type is the multi-spectral image, in Fig. 5, only the reconstructed RGB color images are shown for simplicity. The multi-spectral data for all the images are included in the Supplementary Information [32]. We speculate that the improved resolution is because our image reconstruction process is analogous to fitting the sensor image to a linear combination of structured point-spread functions. Such structured PSFs have been shown to achieve better resolution in localization microscopy [33]. Nevertheless, additional work is needed to convincingly show the mechanism for the improved spatial resolution. The iPhone 6 screen used in the experiment is 326ppi with a pixel size of 78µm. This is much smaller than the resolution of our camera, 4.2cycles/mm and the reconstruction grid, 120µm.

The spatial resolution is on par with some alternative snapshot MSI such as image mapping spectrometry (IMS) [34], imaging spectrometry using a filter stack (IS-FS) [35] and image-replicating imaging spectrometry (IRIS) [36], which provide diffraction-limited resolution. However, other methods have compromised spatial resolutions dictated by the elemental imaging optic, such as imaging spectrometry using a fiber-bundle (IS-FB) [37], imaging spectrometry using a light-field architecture (IS-LFA) [38] and integral field image using hyperpixels [39]. Additionally, the spectral resolution of 9.6nm (Fig. 2(b)) is similar with those of most other snapshot multispectral imagers: IMS [34], IS-FS [37], computed tomography imaging spectrometry (CTIS) [40] and coded aperture snapshot spectral imaging (CASSI) [13].

As discussed in our previous work in [19], spectral resolution is enhanced with respect to smaller pixel size of the diffractive filter, since the diffraction patterns de-correlate more rapidly with the change in wavelength. In addition, both spatial and spectral resolutions are improved with a longer distance $d$ between the DF and the sensor, again due to faster de-correlation. This is in contrast with [19], in which spatial resolution gets worse with larger $d$, since spatial overlaps of the diffraction patterns of neighboring pixels are no longer a limiting factor in the current implementation. Nevertheless, both smaller pixel size and greater $d$ leads to wider spatial spread of the diffraction patterns, which adds to computation burden in reconstruction. Note that the structure profile of the DF is not critical. A number of different computer-generated random profiles are simulated, showing trivial differences in spatial and spectral correlation curves.

We are also able to computationally trade-off spectral resolution against the field of view without changing the hardware. In other words, by calibrating the camera with a larger field of view and lower spectral resolution, we are able to reconstruct images with larger field of view and lower spectral resolution, and vice-versa. To illustrate this, using the same camera and the same image frame (size = 150×150 pixels) as used for the corresponding image in Fig. 3, we reconstructed images of size 50×50 pixels (6.0mm×6.0mm field-of-view) for 9 spectral bands. Some results are plotted in Fig. 6 (see section 12 in the Supplementary Information [32] for details). Thus, our computational approach allows for great flexibility in the selection of resolution in spectrum and space.

Since our camera is based upon computational reconstruction, noise introduced in the reconstruction process has to be clarified. We studied the impact of noise by performing careful simulations, the results of which are summarized in Fig. 5(c). Specifically, we manually added random noise (from a Gaussian distribution) of different standard deviations to a numerically synthesized sensor image, $\mathbf{I}(x',y')$, which is generated using the calibrated PSFs, $\mathbf{A}$ and the multi-spectral object, $\mathbf{S}(x,y,\lambda)$ via $\mathbf{I}=\mathbf{AS}$. The system is then inverted to compute the multi-spectral image using regularization as before. The multi-spectral image is then converted to the RGB color image. Finally, the error between the reconstructed color image and the object in three color channels, averaged over the entire image, is computed. Based on the curve of error versus signal-to-noise ratio (SNR), an SNR tolerance threshold of ~13dB (or equivalently 12 grayscale values for an 8-bit sensor) is estimated. Here, SNR is defined as 10 times the logarithm of the ratio of the average signal intensity over the standard deviation of the Gaussian noise. A detailed experimental analysis (see Supplementary Information [32]) shows that our camera has SNR on par with that of the reference camera under high-light conditions (>500 lux). It experiences only a <0.4dB SNR reduction when the illumination brightness is dimmed to 80 lux. However, at light level as low as 6 lux, it is >2dB worse than a conventional camera. The noise performance at low-light conditions may be further improved by optimizing the design of the diffractive filter to reduce the spatial spread of the diffracted light. Note that decreased SNR is a common issue in the majority of snapshot MSI. Signal level at each detector pixel is reduced since wavelengths are efficiently separated by diffraction [13,34,40], reflection [35] or absorption [14,16-18,38].

Depth-of-field (DOF) is another critical specification of any camera. To experimentally measure the DOF, the multi-spectral PSFs were first captured at various planes of the object (the pinhole) corresponding to various values of defocus (as illustrated in Fig. S12 in the Supplementary Information [32]). Then, the root-mean-squares (RMS) of the differences between the PSF images at a given defocus and those at focus were computed (see Fig. 5(d)). As expected, the RMS increases with larger defocus and a DOF of ±15mm is predicted according to the 13dB threshold (or 12 grayscale, see Fig. 5(c)), which is slightly larger than the ±10mm DOF of the reference camera with the same parameters. The gray plot in Fig. 5(d) is averaged over all the wavelengths. The DOF is shorter in the green and yellow wavelengths compared to that in the blue and red wavelengths.

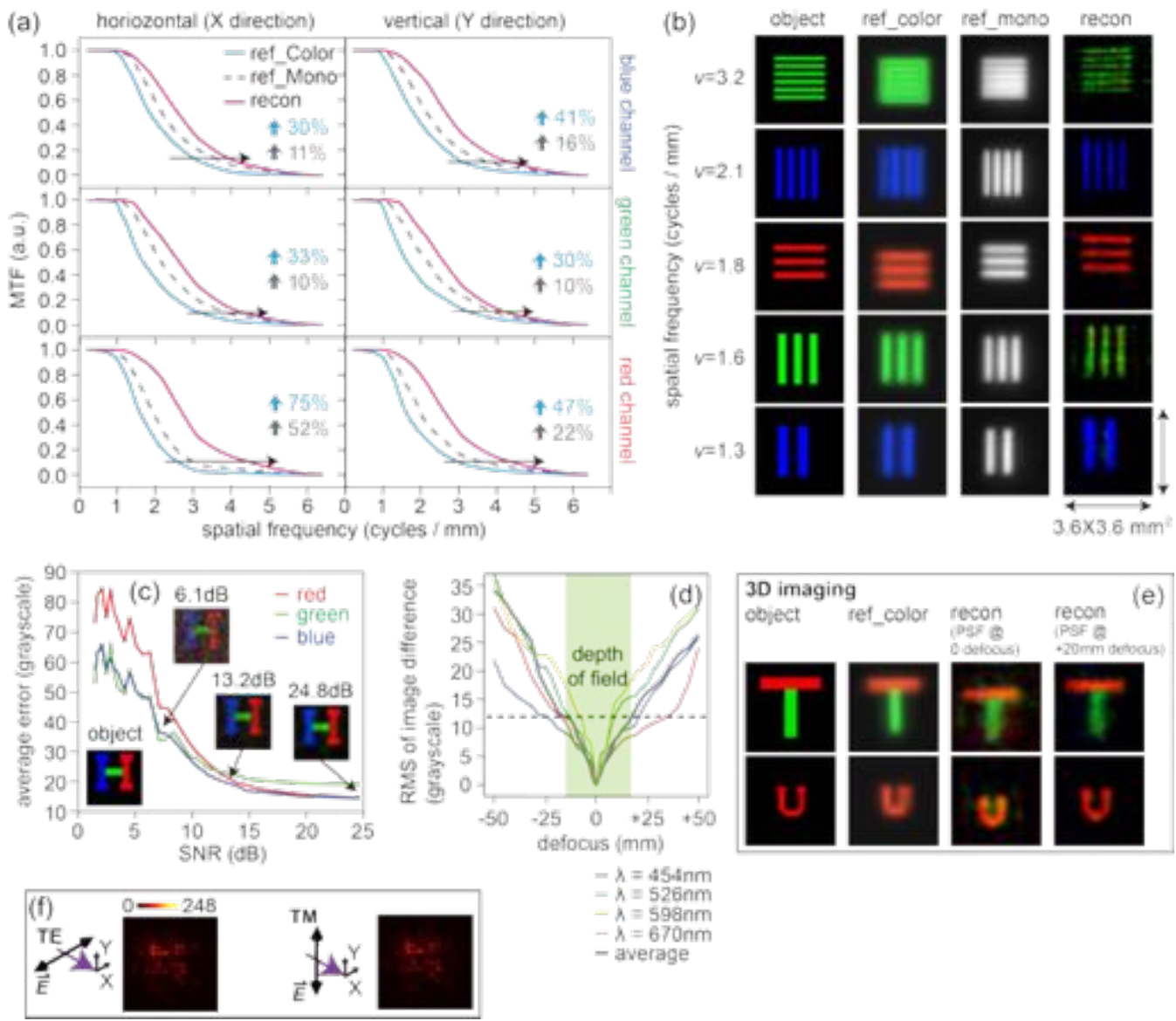

**Fig. 5.** (a) Measured Modulation Transfer Functions (MTFs) in X and Y axes and three primary color channels. (b) Exemplary test object patterns at five spatial frequencies, their reference images using color and monochrome cameras and the reconstruction results. (c) Averaged image reconstruction error versus signal-to-noise ratio (SNR) in three basic color channels. Gaussian noise is added to the numerically synthesized sensor image to change the SNR. (d) Root mean squares (RMSs) of differences between the PSF images with and without defocus. A depth-of-field is estimated (light green region). Blue, green, yellow and red curves are from four example wavelengths. The gray curve is averaged over all wavelengths. (e) Two examples for 3D imaging experiments: the 2-color letter 'T' and the 1-color letter 'U'. (f) Magnified views of the experimentally calibrated PSFs illuminated by two polarization states. TE and TM polarizations are defined on the left side. They are of the same object point and the same wavelength. They are both 70 by 70 sensor pixels.

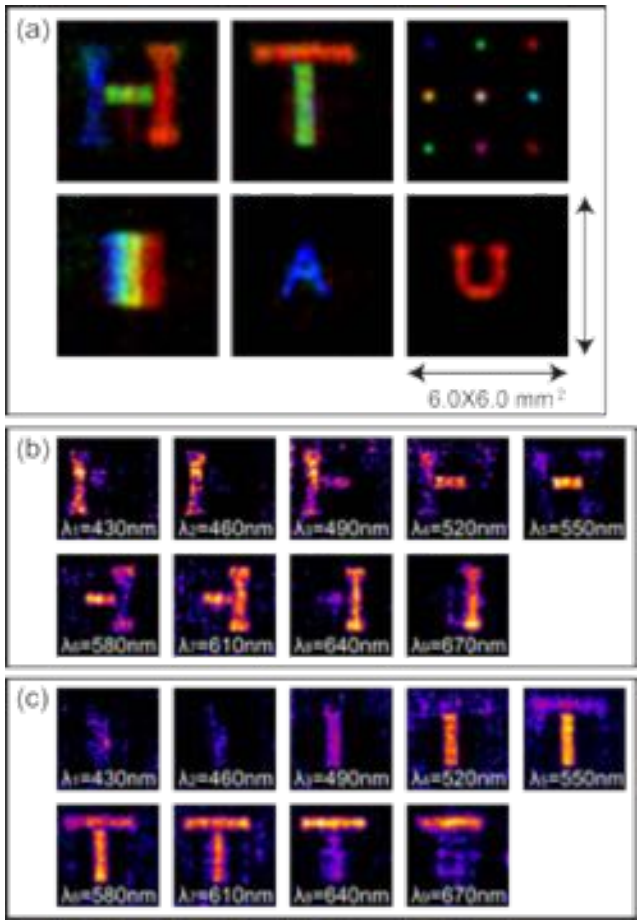

**Fig. 6.** Reconstruction results to demonstrate trade-off between spectral resolution and field-of-view. (a) Synthesized RGB color image of 6×6mm² field-of-view. (b) and (c) Multispectral data of two examples: letter 'H' and letter 'T', respectively. Nine wavelengths from 430nm to 670nm with 30nm spacing are considered.

We also show the capability of our camera in computational refocusing and extension to 3D multi-spectral imaging [41]. A simple demonstration is given in Fig. 5(e). We first measured the multi-spectral PSFs at various planes that were displaced from the in-focus object plane, similar to the DOF experiment performed above. Then, a multi-spectral object was imaged at a plane shifted by 20mm from the in-focus object plane. As expected, the multi-spectral image computed using in-focus PSF data is distorted with significant color errors and noise. However, the correct multi-spectral image can be computationally obtained by simply using the PSF data calibrated from the correct defocus plane (+20mm). This points to the possibility of generating 3D multi-spectral images by calibrating the multi-spectral PSF data over the 3D space. However, the caveat is that this requires a lot more data as well as more computation. This is a feature that most other snapshot MSI do not have. Recently, such depth-dependent PSF induced by coded aperture at Fourier plane [42] and diffuser [43] have been successfully exploited for 3D microscopy [42] and 3D photography [43], respectively. However, they were only implemented for monochromatic light without resolving spectrum contents of the objects.

For a general imaging system, it is important that the diffractive filter is polarization insensitive. Many alternative color filters such as those using plasmonics suffer from the major disadvantage of being polarization sensitive [17,18]. Other multi-spectral imaging techniques exploiting polarization properties of light exhibit polarization-dependent photon throughput [36,44,45], which lose up to 50% of photons when imaging a natural scene of random polarization. To illustrate this advantage of our technique, we experimentally captured the PSF at two orthogonal polarizations (TE and TM) and verified that they are indeed very similar to each other. The magnified views of two diffraction patterns of the same multi-spectral object point ($x$=0, $y$=0, $\lambda$=550nm) are shown in Fig. 5(f). Note that even though we utilized randomly polarized light in all our calibration, the light from the iPhone screen is nearly linearly polarized. This further provides support for the fact that our camera is polarization-independent.

Finally, we characterized the dynamic range of our camera. The images were captured across a wide range of exposure times and then reconstructed. The critical exposure time is 17.86ms. The reconstructed images remain acceptable down to 5ms exposure time. Overexposure (>25ms) deteriorates the reconstruction because more pixels are saturated and the assumption of linearity breaks down.

In our preliminary demonstration, the image size is restricted to 3.6mm×3.6mm, since the PSF is only calibrated over 30×30 object points with step size of 120μm. This is not a fundamental limitation of the technology, but was chosen for simplicity and to ensure fast reconstruction times. A simple approach to increase the image size can be demonstrated by using a sparse image comprised of small blocks of color data as illustrated in Fig. 7(a). The blocks are spaced such that we can treat them independently. This constraint can be removed in the future by accounting for the spatial overlap in the reconstruction algorithms as was discussed previously [19]. Then each block (3.6mm×3.6mm) was calibrated and solved individually. The results are given in Fig. 7(b) and only RGB images are shown for simplicity. The corresponding multi-spectral images are included in the Supplementary Information [32]. In any case, by further optimizing the calibration process and the reconstruction algorithms, it will be possible to perform fast multi-spectral full-frame video imaging in the future [46].

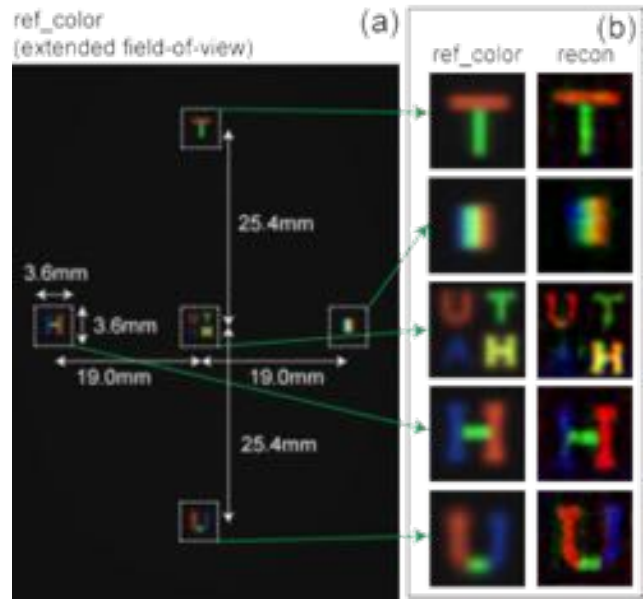

**Fig. 7.** Imaging an extended field-of-view. (a) Reference color image of the whole test pattern over a larger frame. Five small test patterns are included. The field-of-view (FOV) of small patterns and the distance between them are labeled. (b) Reference color images of the individual small patterns and their reconstructed RGB color images.

Imaging beyond the visible light, especially in the near-infrared regime, brings a number of advantages and applications [47,48]. For proof-of-concept, we experimentally calibrated our imaging system at one IR wavelength of 850nm and then conducted visible-IR imaging by simply illuminating the iPhone screen using an 850nm laser beam, as schematically depicted in Fig. 8(a). Figure 8(b) summarizes the reconstruction results of a test pattern of green letter "T". Only a few multi-spectral images are shown for simplicity and the full multi-spectral data is given in the Supplementary Information [32]. Note that the introduction of IR beam causes almost no cross-talk in the visible band. We can clearly discern the shape and the position of the 850nm laser spot in the multi-spectral image. Although the IR spot overlaps with the green pattern, it cannot be identified by the conventional color camera due to the IR cut-filter. More details of the experiments and the cross-talk measurements are included in the Supplementary Information [32].

Since no spatial filter, absorptive filter or polarization component is present in the proposed system, it is able to approach 100% photon throughput. Among all the other snapshot MSI, only IMS [34], IS-FS [35] and CTIS [40] have 100% throughput. CASSI [13] and IRIS [36] have 50% throughput due to binary codes and polarization, respectively. Both IS-LSA [38] and tilted filter array [14-18] have 1/N throughput, where N is the number of spectral bands.

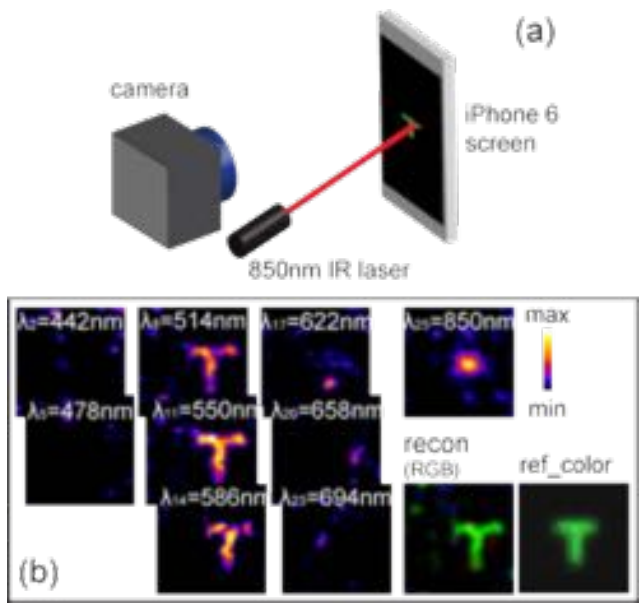

**Fig. 8.** Visible-IR imaging. (a) Schematic of the visible-IR imaging experiment. An 850nm laser illuminates the iPhone 6 screen. Both the DF-based camera and the reference color camera are used to take images. (b) Examples of the reconstructed multi-spectral images. A test pattern of green letter "T" is considered. The 850nm IR spot does not show up in the reference RGB image but can be resolved from the reconstructed multi-spectral data. The RGB color image reconstructed from multi-spectral data within the visible band is also shown.

## 5. CONCLUSIONS

In conclusion, we demonstrated computational multi-spectral video imaging that preserves both spectral and spatial resolutions by simply placing a diffractive filter atop the conventional sensor array, and applying linear reconstruction algorithms. The system exhibits a spatially and spectrally variant PSF, where each multi-spectral object point $(x, y, \lambda)$ is mapped to a set of sensor pixels $(x', y')$. This one-to-many mapping can be inverted via regularization, since it is a linear-transfer function. The inversion process allows us to compute the multi-spectral image. We experimentally demonstrated spectral resolution of 9.6nm and spatial resolution of 4.2cycles/mm, which is higher than that can be achieved with the conventional camera. Since our diffractive filter does not absorb any light, the sensor utilizes all incoming photons. This is in contrast to the conventional color camera, where on an average, 2/3rds of the light is unused due to absorption in the Bayer filter. By removing the Bayer filter, we further make the CMOS sensor fully compatible with the silicon fabrication process. The diffractive filter can be inexpensively replicated via roll-to-roll nano-imprint lithography for mass production [49]. We also raise the possibility of computational 3D multi-spectral imaging with the extension to the computational acquisition of the multi-spectral light-field [50]. Finally, we reiterate that our technology is equally applicable to any portion of the electromagnetic regime as long as the sensor demonstrates sufficient sensitivity. Specifically, we demonstrated the potential for a single sensor to achieve visible-IR imaging.

We should also note that the idea of wavelength encoding has been applied also to achieving super-resolution in space [51-53]. In these approaches, the spectral information is used to encode spatial information with higher resolution. Spectral dispersion combined with computational methods have also been applied to achieve extended depth of focus imaging [54] as well as to reduce aberrations in imaging systems.

**Funding Information.** DOE Sunshot Grant (EE0005959); NASA Early Stage Innovation Grant (NNX14AB13G); Office of Naval Research (N66001-10-1-4065).

**Acknowledgment**. The authors would like to thank Dr. Eyal Shafran for assistance in compiling the multi-spectral video. We also acknowledge the NKT Photonics for assistance with the super-continuum source and the tunable bandpass filter.

See Supplement Information [32] and Supplementary Visualizations 1-4 for supporting content.

**Supplementary Information**

# Computational Multi-Spectral Video Imaging

*Peng Wang[1,&] and Rajesh Menon[1]*

[1] Department of Electrical and Computer Engineering, University of Utah, Salt Lake City, UT 84112, U.S.A.

[&] Current address: Department of Medical Engineering, California Institute of Technology, Pasadena, CA 91125, U.S.A.

# 1.    Device Fabrication

The pixelated multi-level microstructures in the diffractive filter (DF) is patterned by a commercial gray-scale lithography tool. In conventional lithography, a binary mask defines only the transparent and opaque regions (1 and 0 states) for exposure. Therefore, only 2D binary features are generated. On the contrary, in gray-scale lithography, the exposure dose at each point is modulated with different gray-scales [1-3]. In this work, we utilized a Heidelberg microPG101 machine, which works in the direct-laser-writing (DLW) mode where the write head (a focused 404nm solid-state laser) scans through the sample surface pixel by pixel.

Although only two states (exposed and un-exposed) are considered in conventional binary patterning, a photoresist is a non-linear recording media in reality, characterized by a contrast curve. Different depths in accord with different exposure doses are achieved after development. This occurs within a certain range of exposure dose. Take a positive photoresist for example, higher dose leads to deeper feature. The Shipley 1813 photoresist (positive) is used in this work to fabricate the DF. Before patterning structures, it is required to calibrate this contrast curve. Figure S1(a) gives a typical calibration curve. Based on this plot, we can easily derive the exposure dose needed to achieve certain depth of structure. The surface roughness measurement (max-min height ~14.7nm) by atomic-force-microscopy (AFM) demonstrates the excellent smoothness of surface after gray-scale fabrication (Fig. S1(b)).

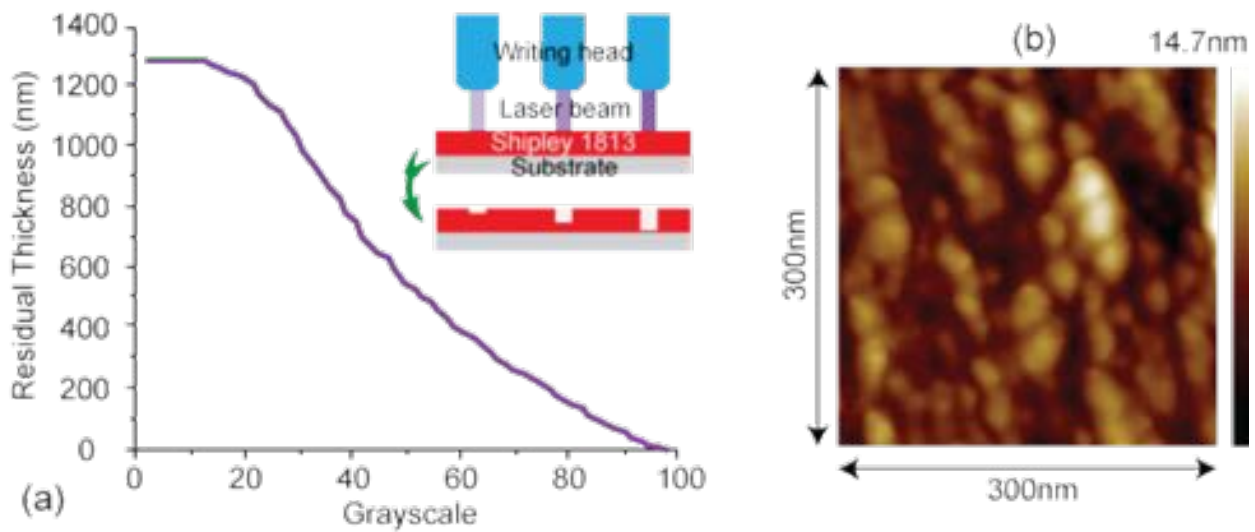

**Figure S1 | (a) Calibration curve of the Shipley 1813 photoresist at laser power of 12mW and duration factor of 40%. Inset: schematic illustration of gray-scale lithography on a positive photoresist. (b) Surface roughness measurement of the fabricated DF by AFM.**

Fabrication procedure:

1) RCA clean a 2-inch fused-silica wafer of 200μm-thick.

2) Spin coat Shipley 1813 photoresist (positive), 45 seconds @ 4500 rpm.

3) Soft bake on a hotplate, 90 seconds @ 115$^o$C.

4) Dehydration in fume hood for ~24 hours.

5) Gray-scale exposure by the Heidelberg microPG101 [3] with 3μm-mode. Power: 12mW; duration factor: 40 %; mode: uni-directional; scan-scale: 2.74MHz.

6) Development in AZ MIF 300 solution for 1 minute.

7) Rinse in DI water for 1 minute. Dry the sample by blowing Nitrogen.

## 2.    Setups

### 2.1 Calibration setup

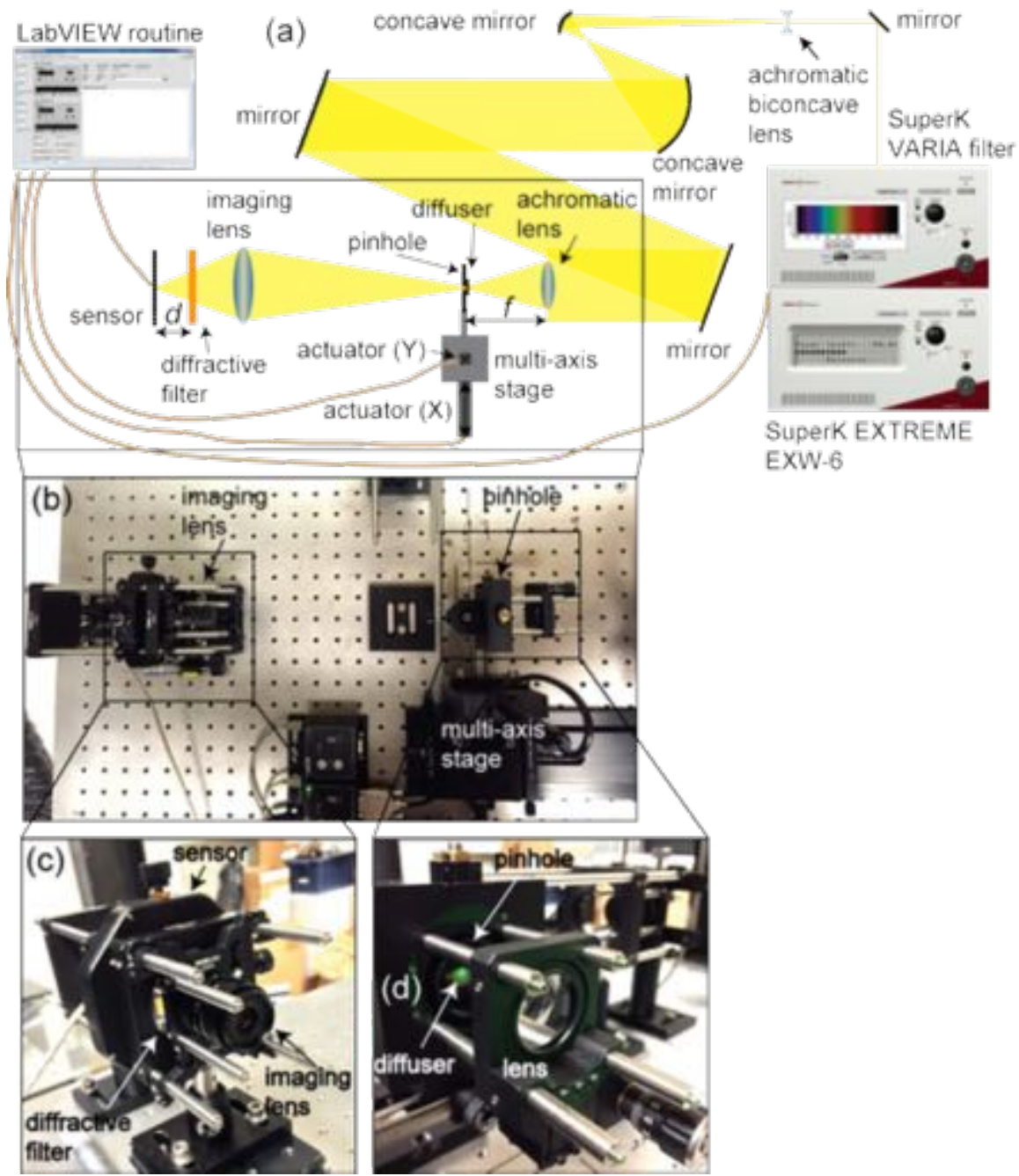

**Figure S2 | (a)** Schematic of the setup measuring all the PSFs in the system matrix A for the visible band. **(b)** Photograph of the calibration part of the setup (black box in (a)). Close-up views of the DF-sensor assembly **(c)** and the pinhole-assembly **(d)**, enclosed by the small black boxes in (b).

The calibration system is schematically shown in Fig. S2(a). It is used to measure the point-spread-functions (PSF) of the constructed camera of DF-sensor assembly in the visible band. The broadband laser light from the super-continuum source is pre-conditioned by a set of mirrors and an achromatic biconcave lens to expand and collimate the beam. In the calibration part, enclosed by the black box, an achromatic lens (AC254-100-A, $f$=100mm, Thorlabs) focuses light onto the 150μm-diameter pinhole with a piece of diffuser glued on its backside. The pinhole, together with the achromatic lens, is mounted onto a two-axis motor stage (Z825B, Thorlabs), which scans the object plane in both X and Y directions. The stepping resolution is 0.1μm. Then, the imaging lens (MVL12WA, $f$=12mm, Thorlabs), the diffractive filter (DF) and the monochrome CMOS sensor (DMM 22BUC03-ML, The Imaging Source) are assembled to make the multi-spectral imager. The DF and the sensor are separated by a gap $d$. The super-continuum source (SuperK EXTREME EXW-6, NKT Photonics) is equipped with a tunable bandpass filter (VARIA, NKT Photonics), which selects the wavelengths from 430nm to 718nm with 12nm spacing and 12nm bandwidth. Note that the super-continuum source, the motor stage and the sensor are all controlled and synchronized via a custom-built LabView program. The photographs of calibration setup are shown in Figs. S2(b) – (d). In the current setup, an entire calibration (30 by 30 points and 25 wavelengths) takes at least 3 hours to complete.

## 2.2 <u>Imaging setup</u>

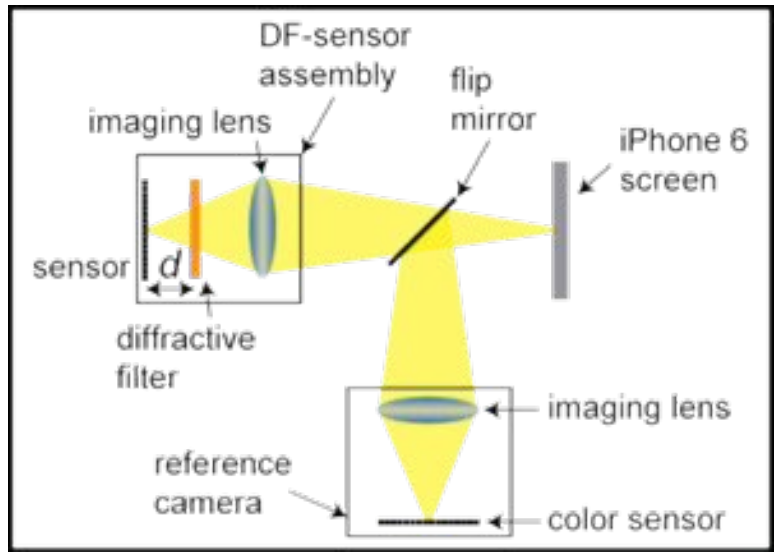

**Figure S3 | Schematic of the imaging setup in the visible band. An iPhone 6 screen displays the test patterns and a flip mirror switches between the studied multi-spectral camera and the reference camera.**

The setup for multi-spectral imaging in the visible band is schematically illustrated in Fig. S3. As mentioned in the main text, an iPhone 6 screen is used to display all the test patterns. The multi-spectral imager of DF-sensor assembly is the same with the one in the previous calibration experiment. A large flip mirror is inserted in the optical path to switch between the DF-sensor assembly and the reference camera. The reference arm is made of the same imagine lens and the same sensor chip but with Bayer filter array on top (a color sensor, DFM 22BUC03-ML, The Imaging Source).

The gain of the sensor is set to its minimum value to avoid background noises. Both binning and noise reduction (averaging over a number of frames) functionalities are disabled. The default de-Bayering technique 'edge sensing' is selected. In addition, auto white balance is used in the reference color sensor. For the objects used here, the white balance values are typically 70, 64 and 68 for the red, green and blue channels, respectively.

## 2.3 Calibration setup (visible-IR)

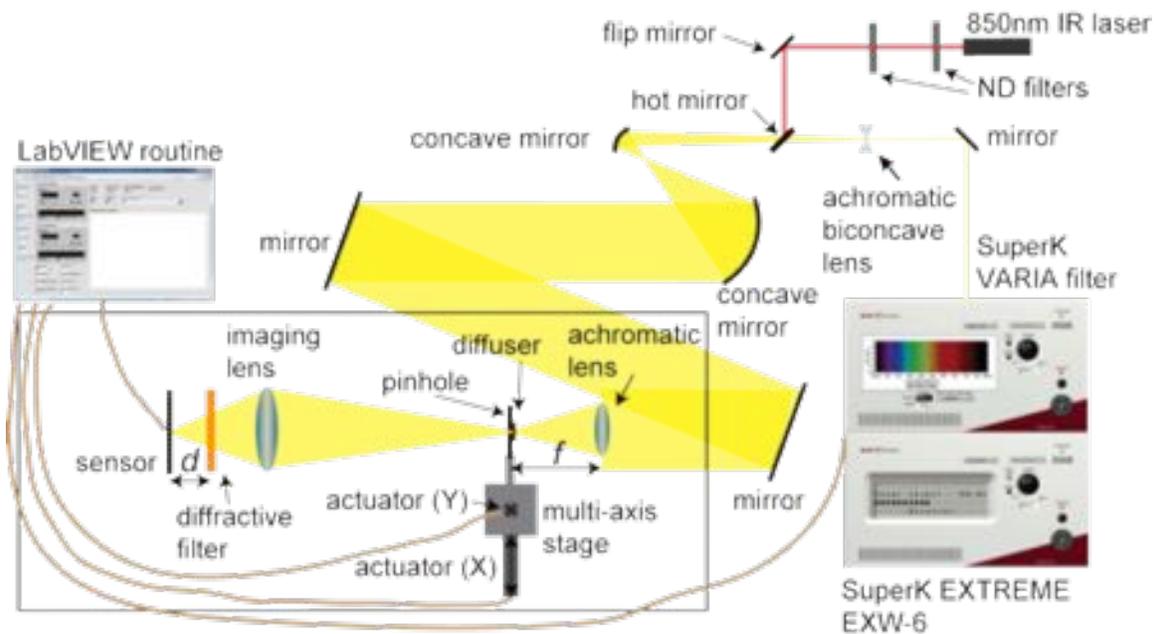

**Figure S4 | Schematic of the setup measuring all the PSFs in the system matrix A for the visible-IR band. An 850nm IR laser is combined with the light from the super-continuum source through a hot mirror.**

Figure S4 gives the schematic of the calibration setup for the visible-IR imaging. It is the same with the calibration setup for visible light imaging in Fig. S2(a), except that an 850nm IR laser is introduced and combined with the visible light from the super-continuum source through a hot mirror. Two ND filters and a mirror are used to pre-condition the IR beam. The same LabView routine controls the system. Still 25 wavelengths in total are considered in calibration: the visible band from 430nm to 706nm with 12nm spacing (24 wavelengths) and the IR light (850nm).

## 2.4 Imaging setup (visible-IR)

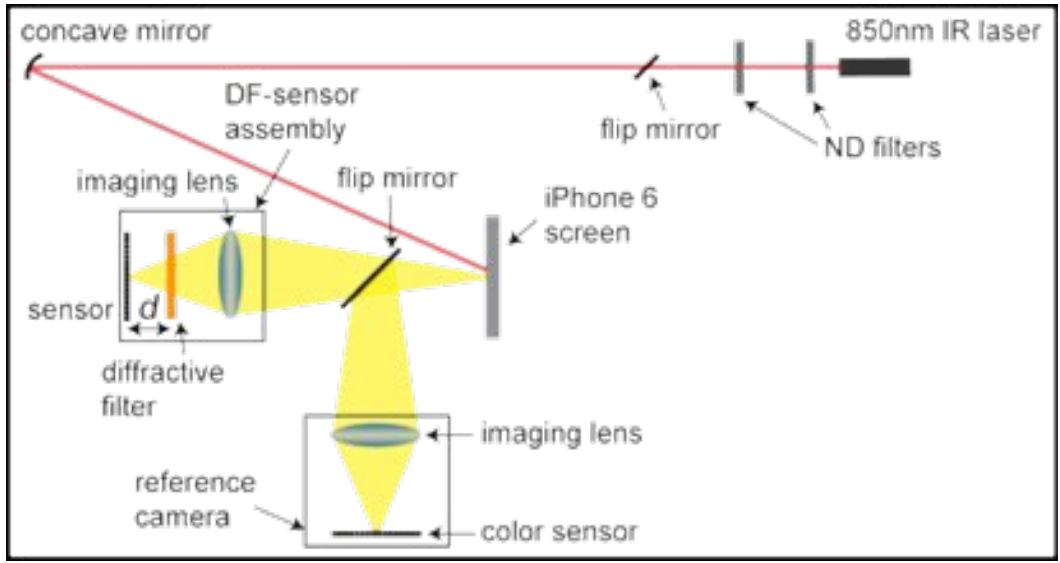

**Figure S5 | Schematic of the imaging setup in the visible-IR band. An 850nm IR laser beam is directed to shoot the iPhone 6 screen, overlapping with the test pattern on the screen.**

The visible-IR imaging setup is schematically drawn in Fig. S5. It is the same with the imaging setup for the visible light in Fig. S3, except that an 850nm IR laser is introduced. This laser beam is pre-conditioned as in the calibration system (Fig. S4) and then guided towards the iPhone 6 screen. The beam spot overlaps with the test patterns displayed on the screen. Its spot size is reduced by the concave mirror.

# 3. Correlation Functions

Cross-correlation evaluates how similar two diffraction patterns shifted in either spatial or spectral domains are [4-7]. It can be generally written as:

$$C(x, y, \Delta\lambda) = \left\langle \int \mathbf{A}(x', y'; x, y, \lambda) \cdot \mathbf{A}(x', y'; x, y, \lambda + \Delta\lambda) dx' dy' \right\rangle, \qquad \text{(S1.1)}$$

$$C(\Delta x, y, \lambda) = \left\langle \int \mathbf{A}(x', y'; x, y, \lambda) \cdot \mathbf{A}(x', y'; x + \Delta x, y, \lambda) dx' dy' \right\rangle, \qquad \text{(S1.2)}$$

$$C(x, \Delta y, \lambda) = \left\langle \int \mathbf{A}(x', y'; x, y, \lambda) \cdot \mathbf{A}(x', y'; x, y + \Delta y, \lambda) dx' dy' \right\rangle. \qquad \text{(S1.3)}$$

Inside the integral over the image space ($x'y'$) is a product of the diffraction patterns separated by a certain distance in either spectral ($\Delta\lambda$) or spatial ($\Delta x$ and $\Delta y$) domains. The bracket <…> represents average over either all the wavelengths (Eq. S1.1) or all the positions (Eqs. S1.2 and S1.3). Usually, the correlation function $C$ drops with increasing shifts ($\Delta\lambda$, $\Delta x$ and $\Delta y$).

Figures S6(a) and (b) plot the spatial correlation functions at all different wavelengths, in X and Y direction, respectively. The spatial resolution is higher in the blue and green spectra (442nm to 598nm), but is lower in the 430nm and the red regime (> 610nm). This behaves the same for both X and Y directions. Figure S6(c) plots the spectral correlation functions at some exemplary spatial locations. For clarity, they are manually offset a certain gap. As we can see, the spectral resolution does not change too much over the entire calibration space or the field-of-view (3.6mm by 3.6mm).

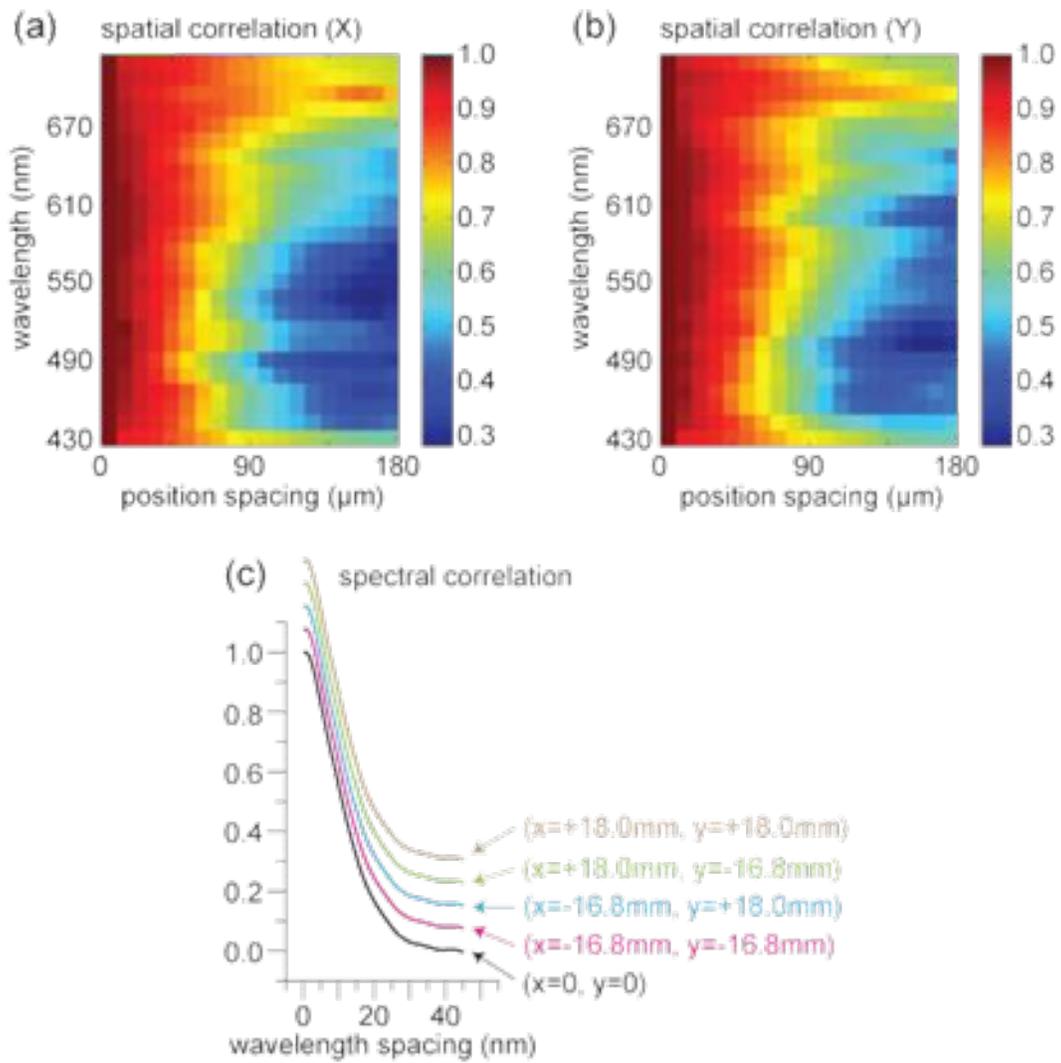

**Figure S6 | Measured spatial correlation functions at all different wavelengths in X (a) and Y (b) directions. (c) Measured spectral correlation functions at some exemplary spatial positions.**

# 4.    Regularization Algorithm

## 4.1 Vectorization

In the matrix representation of the imaging problem **I**=**AS**, we need to vectorize the components for the ease of computation. **I** is the 2D monochrome image, it is vectorized into a single column. It has 150×150=22500 elements. **S** is the 3D multi-spectral data, it is also vectorized into a single column. It has 30×30×25=22500 elements, as well. Thus, we need to convert the 5-D **A**($x'$,$y'$;$x$,$y$,$\lambda$) into a 22500×22500 matrix. As a result, each column of the matrix **A** is the vectorized version of the diffraction pattern (or PSF) at one wavelength $\lambda$ and one spatial location ($x$,$y$).

## 4.2 Regularization

Regularization is used to solve the inverse problem **S**=**A**$^{-1}$**I**. This is necessary since the system matrix **A** is often ill-conditioned and it's impractical to calculate its direct inverse [8,9]. To start with, **A** is analyzed by singlular-value-decomposition (SVD). It decomposes **A** into a representation of a sequence of singular values and a set of left and right singular vectors: **A** = **UΣV** [8]. **Σ** is a diagonal matrix with the singular values as its diagonal elements, arranged in a descending manner ($\sigma_1 \geq \sigma_2 \geq \sigma_3 \geq \ldots \geq \sigma_k$). The columns of **U** and **V** matrices ($u_1$, $u_2$, $u_3 \ldots u_k$, and $v_1$, $v_2$, $v_3 \ldots v_k$) contain the $k$×1 left and right singular vectors, respectively. SVD is equivalent to a Fourier expansion of the system matrix. The vectors $u_i$ and $v_i$ are the frequency components and $\sigma_i$ are the coefficients of the components ($i$=1,2,3…$k$). From the SVD point-of-view, any forward problem diminishes the high-frequency components in $u_i$ and $v_i$. However, the inverse process attempts to magnify those high-frequency parts. The regularization stabilizes the problem by minimizing both the residual norm ‖**AS-I**‖$_2$ and the solution norm ‖**S**‖$_2$. Usually, a regularization parameter $\omega$ is selected to balance these two terms. Tikhonov regularization is a widely-used method [8]. Mathematically, its goal is stated as:

$$\min\left\{\left\|\mathbf{AS} - \mathbf{I}\right\|_2^2 + \omega^2 \left\|\mathbf{S}\right\|_2^2\right\}. \tag{S2}$$

From the computational perspective, it can be equally formulated as applying filter factors to solution vectors:

$$\mathbf{I}_\omega = \sum_{i=1}^{n} \phi_i^{[\omega]} \frac{u_i^T \mathbf{S}}{\sigma_i} v_i, \tag{S3}$$

in which the filter factor is defined as:

$$\phi_i^{[\omega]} = \frac{\sigma_i^2}{\sigma_i^2 + \omega^2}. \tag{S4}$$

In Eq. (S4), we call $\mathbf{I}_\omega$ the Tikhonov regularization solution with parameter $\omega$.

## 4.3 Regularization parameter

Figure S7(a) plots the singular values $\sigma_i$ of the PSF matrix $\mathbf{A}$ and Fig. S7(b) is a zoom-in plot. It is relatively smooth and flat with large values, indicating potentially good reconstruction results. An appropriate parameter $\omega$ has to be chosen in Tikhonov regularization [9]. An $\omega=3.0$ (Fig. S7(d)) was selected for this work. Figure S7 also gives two more examples. Much too small $\omega=0.5$ (Fig. S7(c)) leads to results overwhelmed by noises, while much too large $\omega=18$ (Fig. S7(e)) leads to over-smoothed distorted results.

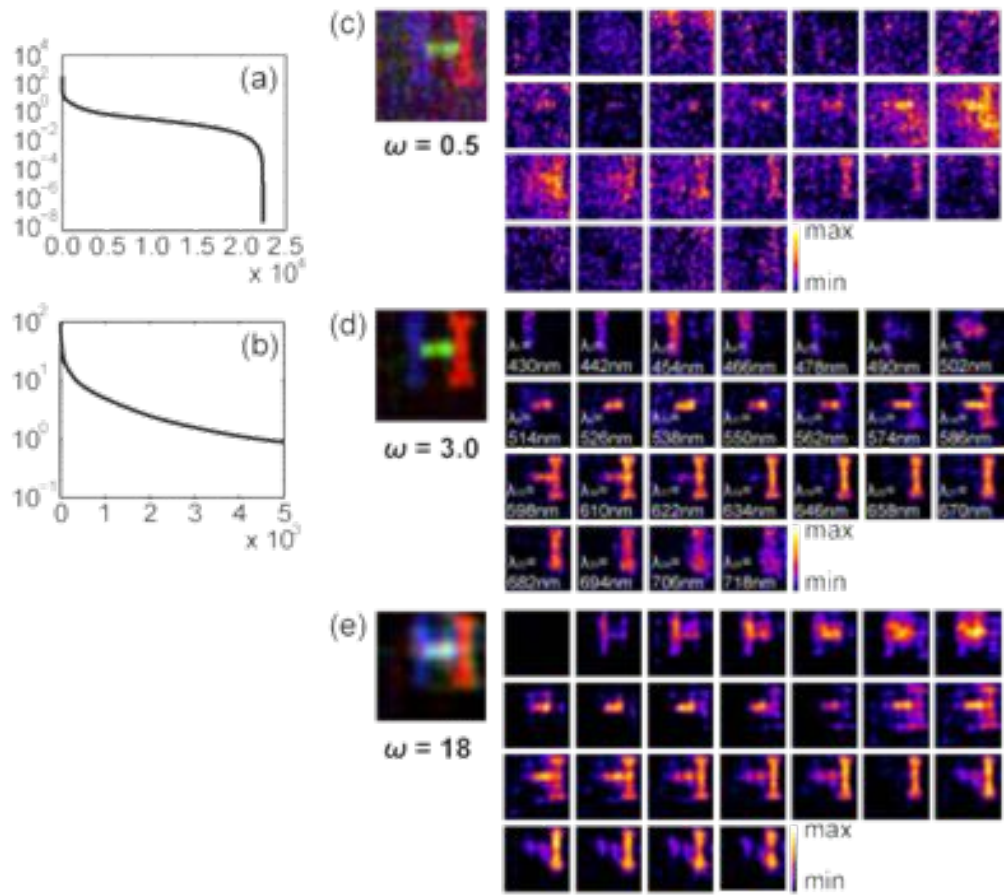

**Figure S7 | (a) Singular values $\sigma_i$ of the PSF matrix A. (b) A zoom-in plot of $\sigma_i$. Reconstruction results using regularization parameters (c) $\omega=0.5$; (d) $\omega=3.0$; (e) $\omega=18$.**

# 5.    Noise Analysis

## 5.1 Noise sensitivity

Figure 4(c) in the main text gives one example on how random noise affects the reconstruction performance. Figure S8 summarizes some more examples of other test patterns. Similarly, a threshold of ~13dB may be estimated. Again, the raw monochrome images are numerically synthesized by multiplying the experimentally measured PSFs with the multi-spectral images of the original test patterns. Then Gaussian noises of various standard deviations are added to the raw images. Finally, regularization is executed to recover the multi-spectral data and the RGB color images. SNR is defined in the main text. Errors averaged over the entire image in three basic color channels are plotted.

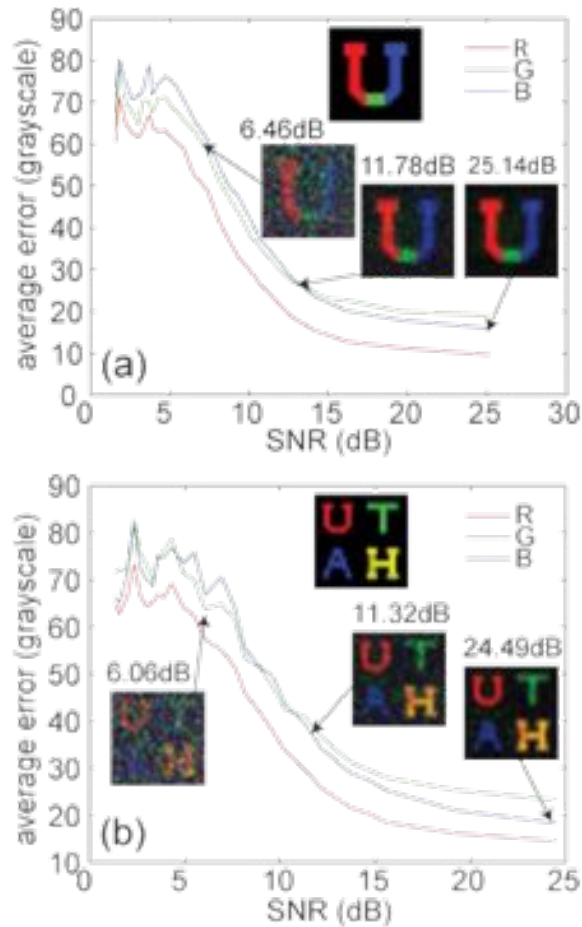

**Figure S8 | Averaged image reconstruction errors versus SNRs in three basic color channels. Gaussian noise is added to the numerically synthesized images. The test patterns are: (a) three-color letter 'U'; (b) four-color letters 'U' 'T' 'A' and 'H'; (c) Mono Lisa; (d) Superman logo.**

## 5.2 Signal-to-noise ratio (SNR)

In order to experimentally determine the signal-to-noise ratio (SNR) of our constructed multi-spectral imager, we devised the following experiments.

1. Use test patterns of uniform color: blue, green, red and gray.

2. Capture raw monochrome images by the constructed camera.

3. Reconstruct the multi-spectral images and the RGB images by regularization.

4. Repeat steps 1 – 3 for 1000 times for each uniform color.

5. Compute the mean and the standard deviation of the reconstruction result of each frame. The mean is the ground truth signal and the standard deviation represents the noise. Calculate the signal-to-noise ratio (SNR) in the logarithm scale for all frames of each uniform color.

6. Plot SNR versus all frames, against SNR of the reference images captured by the conventional RGB camera.

7. Repeat steps 1 – 6 for both high brightness (>500 lux) and low brightness (~6 lux) of iPhone 6 screen. For low brightness 500 frames are considered.

Figures S9 – S11 summarize the SNR plots of both the DF-based multi-spectral camera and the reference color camera for the cases of high brightness, middle brightness and low brightness illuminations, respectively. At high brightness, two exposure times are tested. As can be observed, our multi-spectral imager provides SNR similar to the conventional color camera at high brightness and critical exposure (Figs. S9(a) and (b)). At a moderate light condition (80 lux) and critical exposure, the multi-spectral imager only experiences a <0.4dB degradation in its SNR (see Fig. S10), compared to that at high brightness. This indicates that our imager is able to perform well even when the illumination condition is reduced by more than 7 times.

Note that for the low brightness condition (Fig. S11), an exposure times of 250ms is used for all test patterns. Technically, this is the longest exposure time that the sensor provides. At this exposure time, the monochrome sensor is under exposed, according to the maximum pixel values in the monochrome raw images. This is the primary reason why the SNR experiences significant drop and is worse than the reference color camera. Another sensor with longer exposure time may easily alleviate this limitation. In addition, low-light performance of the imager can also be improved by carefully engineering the diffractive filter such that the spatial spread of the diffraction pattern is minimized.

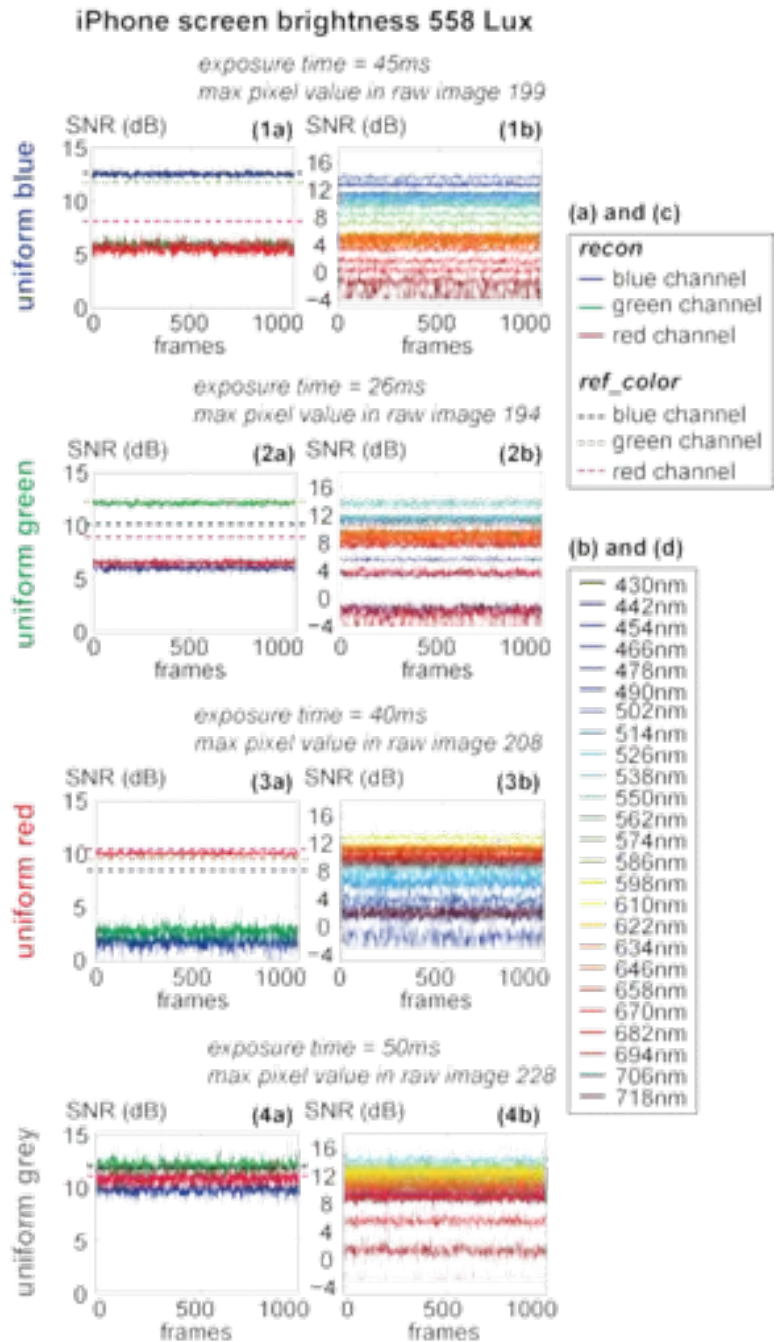

**Figure S9 | Experimentally measured SNRs for 1000 frames at high brightness condition (558 lux). The reconstruction results are in solid lines, while those of the reference color camera are in dashed lines. Four test patterns of uniform colors are considered: (1) blue; (2) green; (3) red; (4) gray. Both RGB images ((a) and (c)) and multi-spectral images ((b) and (d)) are calculated and plotted. Two exposures times are tested: (a) and (b) critical exposure; (c) and (d) under-exposure. The maximum pixel values in the raw monochrome sensor images are also given for reference.**

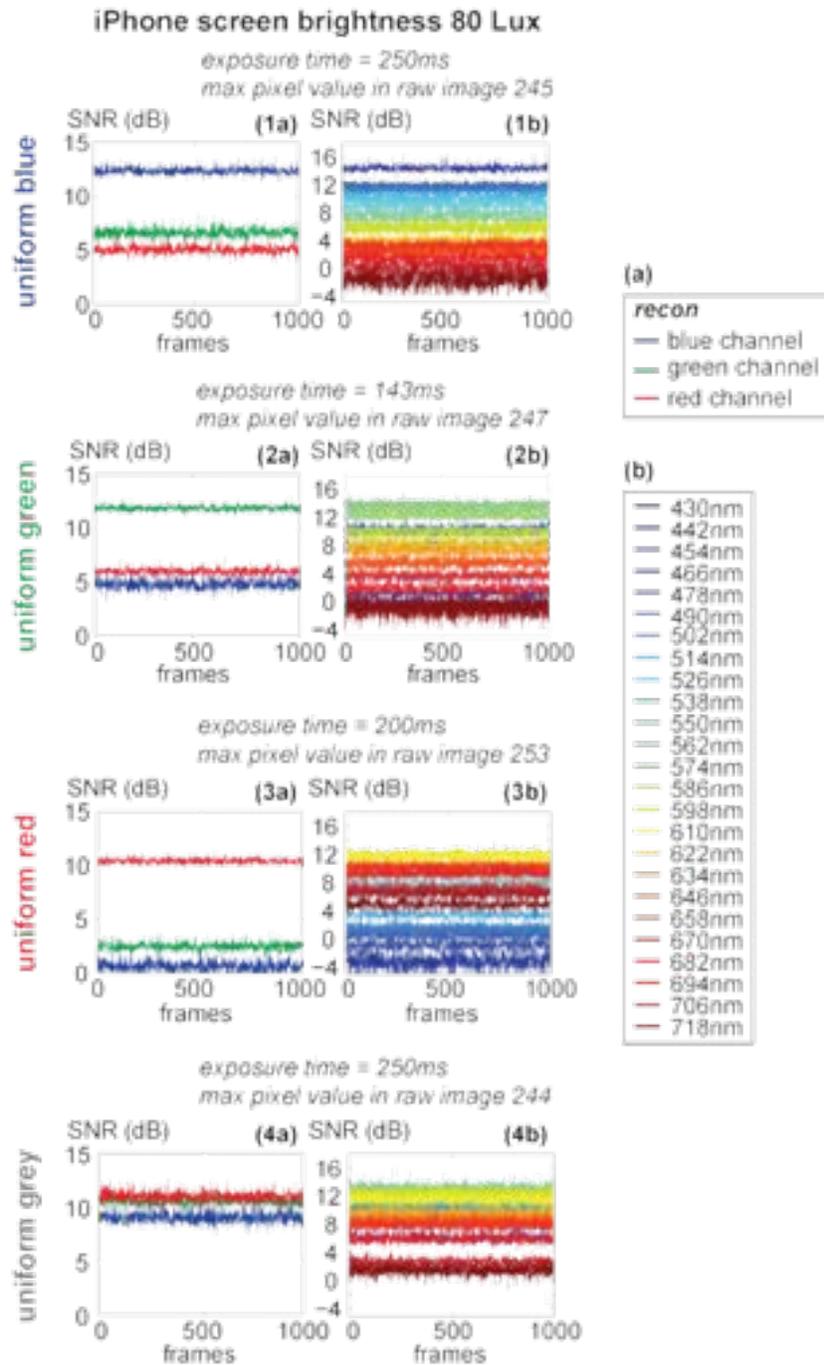

**Figure S10 | Experimentally measured SNRs for 1000 frames at middle brightness condition (~80 lux). The reconstruction results are in solid lines, while those of the reference color camera are in dashed lines. Four test patterns of uniform colors are considered: (1) blue; (2) green; (3) red; (4) gray. Both RGB images (a) and multi-spectral images (b) are calculated and plotted. The exposures times are set to give critical exposures. The maximum pixel values in the raw monochrome sensor images are also given for reference.**

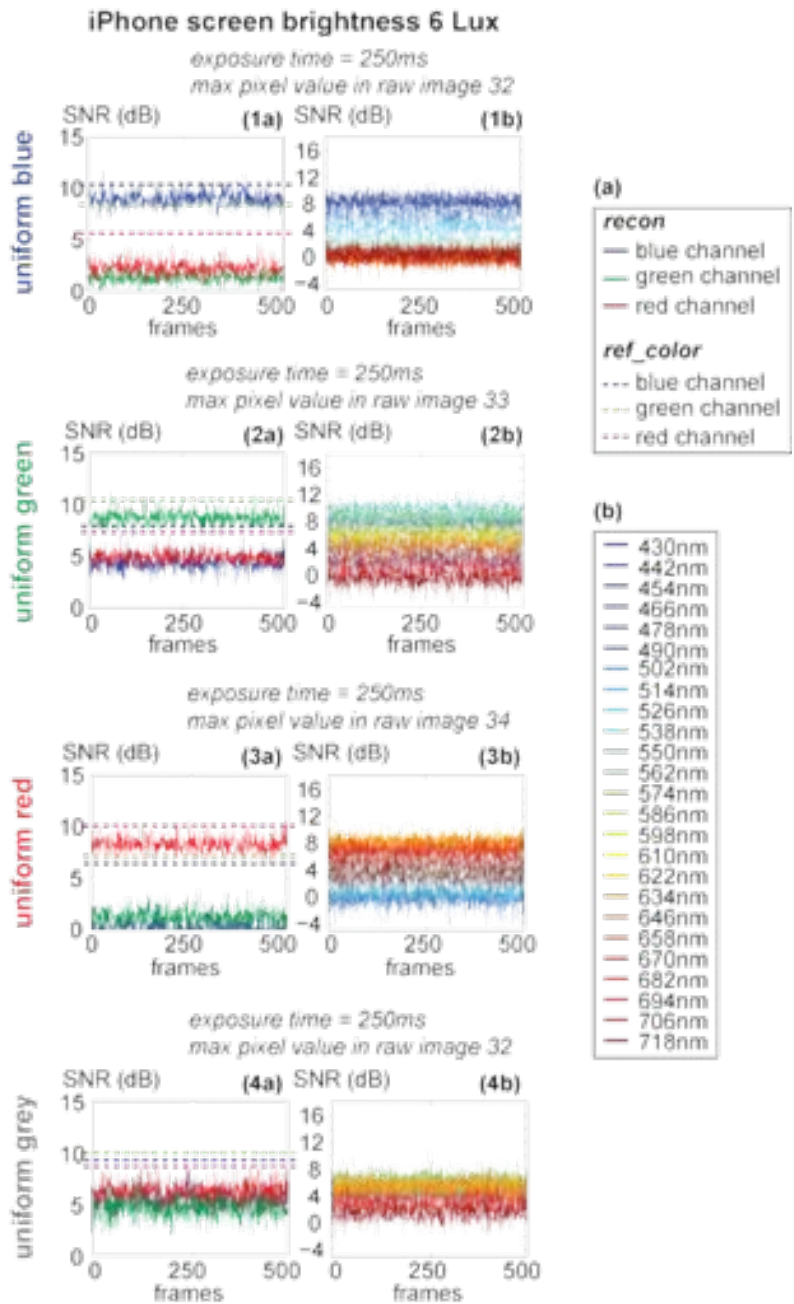

**Figure S11 | Experimentally measured SNRs for 500 frames at low brightness condition (~6 lux). The reconstruction results are in solid lines, while those of the reference color camera are in dashed lines. Four test patterns of uniform colors are considered: (1) blue; (2) green; (3) red; (4) gray. Both RGB images (a) and multi-spectral images (b) are calculated and plotted. The maximum exposure times are used, but still under-exposed. The maximum pixel values in the raw monochrome sensor images are also given for reference.**

# 6.     Spectrum Accuracy

In order to evaluate how accurate the studied multi-spectral imager can recover the spectrum information, we captured raw monochrome images of the single pinhole illuminated by different spectra of light. The spectra are generated by super-continuum source passing through different color filters used in photography (Nikon). The reconstructed spectra, extracted from the multi-spectral data, and the reference spectra measured by a commercial portable spectrometer (Ocean Optics Jaz [10]) are plotted in Fig. S12. Spectra of seven different colors are tested: blue, green, red, brown, orange, yellow and pink. The photographs of the color filters and the raw monochrome images are shown as insets. The reconstructed spectra match quite well with the reference. The root-mean-squares (RMS) of the errors between the reconstructed and the reference spectra are calculated and listed in Table S1. The RMS of error averaged over all seven colors is about 7.2%, which again indicates that our multi-spectral imaging system is able to reconstruct the spectral information with reasonable accuracy.

**Table S1. RMS of the error between the reconstructed and the reference spectra**

| Color filter | RMS of error |
|---|---|
| *Blue* | 8.37% |
| *Green* | 8.92% |
| *Red* | 5.35% |
| *Brown* | 7.52% |
| *Orange* | 6.89% |
| *Yellow* | 6.32% |
| *Pink* | 7.14% |
| Average | 7.21% |

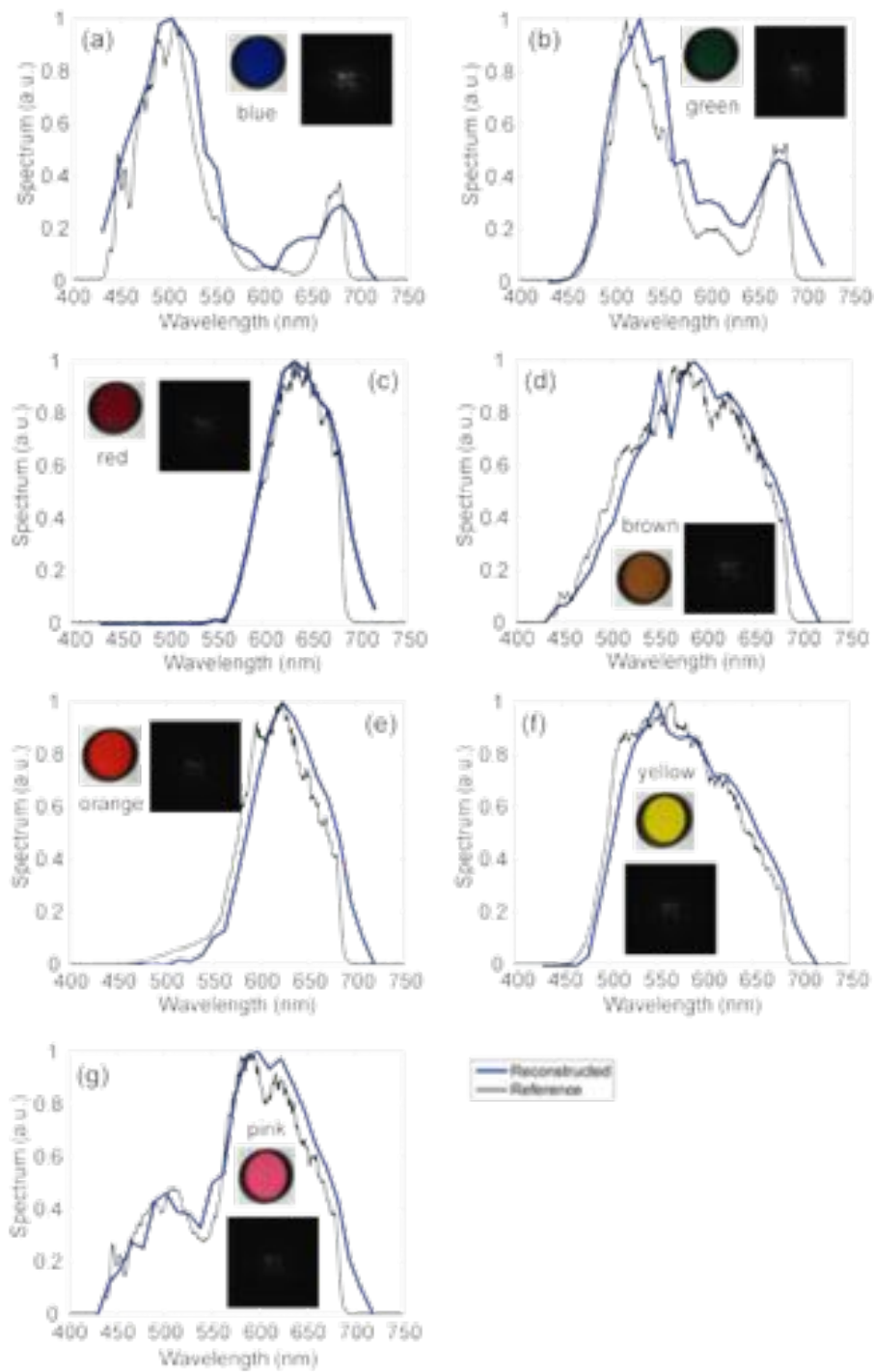

**Figure S12 | Experimentally reconstructed spectra (blue solid lines) and reference spectra measured by a commercial spectrometer (black solid lines). The pinhole is illuminated by the super-continuum source passing through: (a) a blue filter; (b) a green filter; (c) a red filter; (d) a brown filter; (e) an orange filter; (f) a yellow filter; (g) a pink filter. Insets: the photographs of the color filters and the raw monochrome images.**

# 7.    Depth-of-field

To study the depth-of-field (DOF) of the multi-spectral imager, we experimentally calibrated the PSFs at various defocus positions. Figure S13 summarizes some exemplary PSFs measured at nine different defocus positions and six different wavelengths. They are all at the same position $x$=0, $y$=0. As expected, for all the wavelengths, the PSFs changes gradually along the defocus axis (z). Therefore, when the test pattern is placed at certain distance away from the calibrated object plane, prominent errors may occur in the reconstruction.

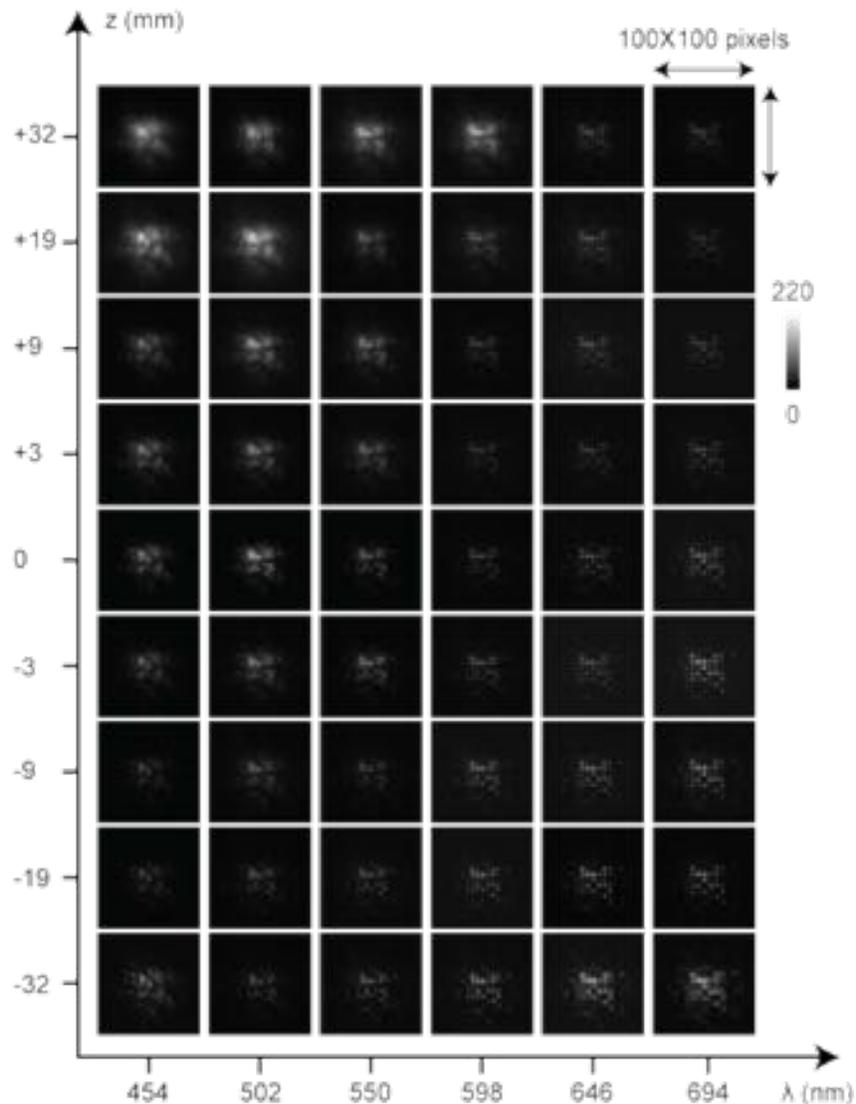

**Figure S13 | Experimentally calibrated PSFs at various defocus positions (z), various wavelengths (λ) and fixed object position (x=0, y=0). Nine defocus positions and six wavelengths are plotted for simplicity. Each PSF image is cropped to 100 by 100 sensor pixels from the original images.**

In order to understand how much is the system tolerant to defocus, it is necessary to calculate the DOF. This is obtained by computing the RMS of the difference between the PSFs at defocus and at focus (z=0) versus defocus values. The data averaged all wavelengths is plotted in Fig. 4(d) in the main text. Here, Fig. S14 gives the plots at some exemplary wavelengths. As we can see, green and yellow wavelengths have smaller DOF compared to those of blue and red wavelengths.

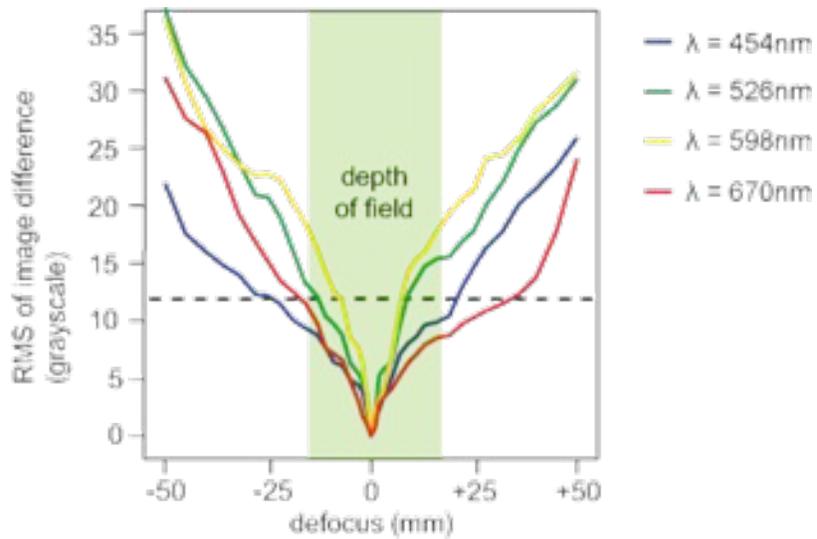

**Figure S14 | Experimentally measured RMS of the difference of the PSF images between at defocus and at focus (z=0). Four wavelengths are plotted as examples: 454nm (blue); 526nm (green); 598nm (yellow) and 670nm (red).**

# 8. Other Multi-spectral Images

## 8.1 Multi-spectral images in MTF measurements

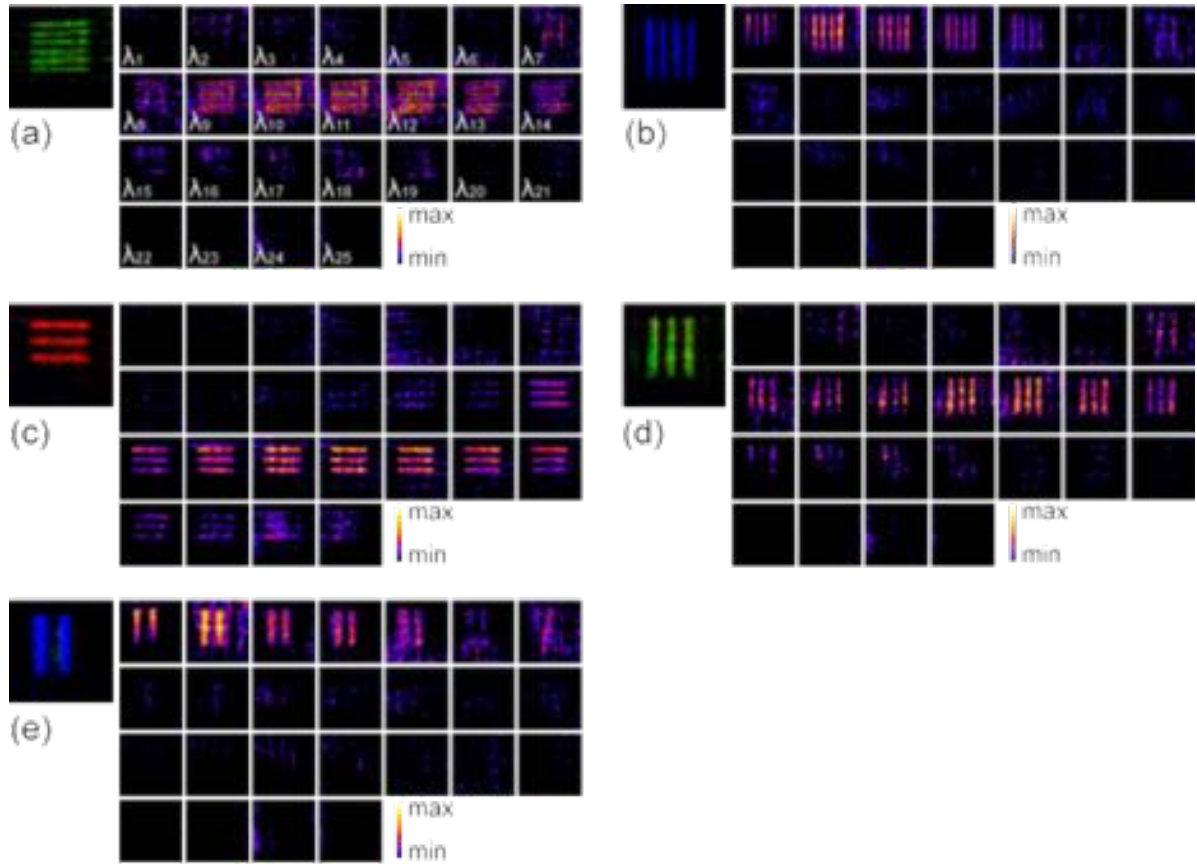

**Figure S15 | Experimentally reconstructed multi-spectral images of the test patterns for MTF measurements:** (a) horizontal green lines at 3.2 cycles/mm frequency; (b) vertical blue lines at 2.1 cycles/mm frequency; (c) horizontal red lines at 1.8 cycles/mm frequency; (d) vertical green lines at 1.6 cycles/mm frequency; (e) vertical blue lines at 1.3 cycles/mm frequency.

## 8.2 Multi-spectral images in 3D imaging

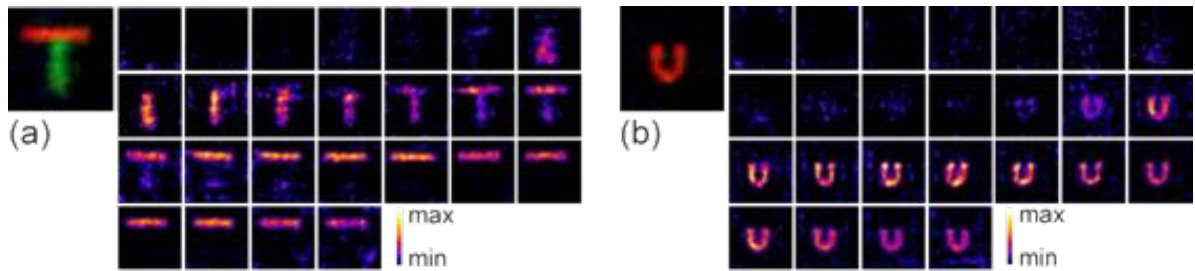

**Figure S16 | Experimentally reconstructed multi-spectral images of the test patterns for 3D imaging: (a) 2-color letter 'T'; (b) 1-color letter 'U'. The object is placed at +20mm defocus and the PSFs is also calibrated at +20mm defocus.**

## 8.3 Multi-spectral images in measuring dynamic-range

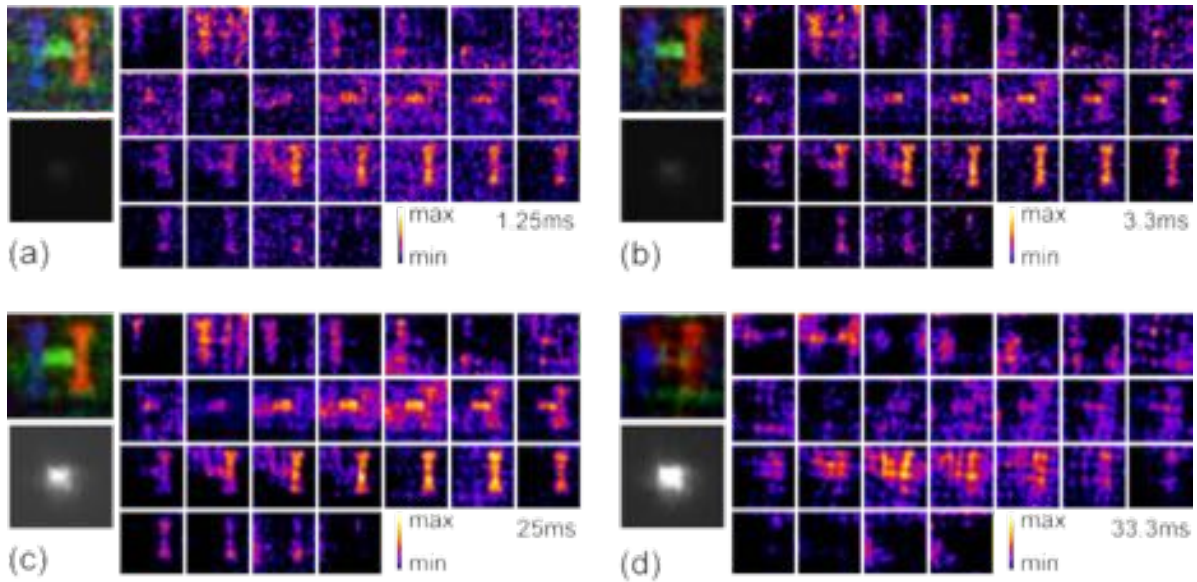

**Figure S17 | Experimentally reconstructed multi-spectral images of the test patterns for measuring the dynamic-range of the multi-spectral imager. The exposure times are: (a) 1.25ms; (b) 3.3ms; (c) 25ms; (d) 33.3ms. The corresponding raw monochrome images are also shown in grayscale.**

## 8.4 Multi-spectral images in imaging the extended field-of-view

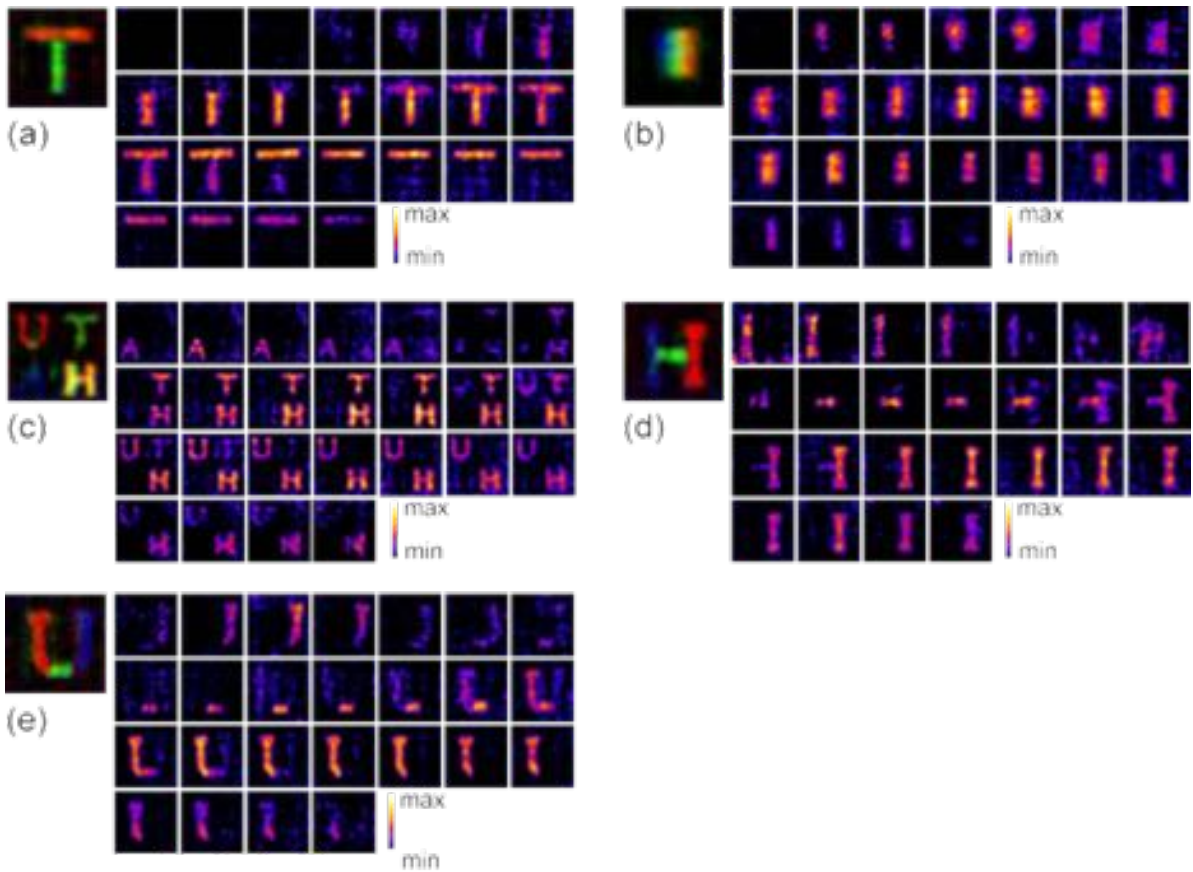

**Figure S18 | Experimentally reconstructed multi-spectral images of the test patterns for imaging the extended field-of-view. (a)** 2-color letter 'T' at the top; **(b)** rainbow on the right; **(c)** 4-color letters 'U' 'T' 'A' 'H' in the center; **(d)** 3-color letter 'H' on the left; **(e)** 3-color letter 'U' at the bottom.

# 9. Non-black Background Imaging

All the object patterns tested in this work are in black background. Therefore, it is necessary to understand how the multi-spectral imager performs when the background is not black, but in grayscale. Figures S19(1), (2) and (3) summarize the experimental results when the background is in three grayscale levels: 40, 80 and 127, respectively. The original patterns to be displayed on iPhone screen, the raw monochrome images and the reconstruction outputs are shown. Note that the references are uniform of grayscales the same with those of the test patterns. The white boxes in (a) and (b) represent the area of object plane which is calibrated (3.6mm by 3.6mm). If a constant regularization parameter $\omega$=3.0 is used, which is the same with all the multi-spectral imaging experiments done in this work, we can observe that significant image quality degradation with brighter background grayscale (see Figs. S19(1e), (2e) and (3e)). However, if an optimal regularization parameter $\omega$ is applied for each situation of different background grayscale, the quality of reconstruction degrade much slower (see Figs. S19(1f), (2f) and (3f)). Here, $\omega$=6, $\omega$=10 and $\omega$=35 are chosen for background grayscales of 40, 80 and 127, respectively.

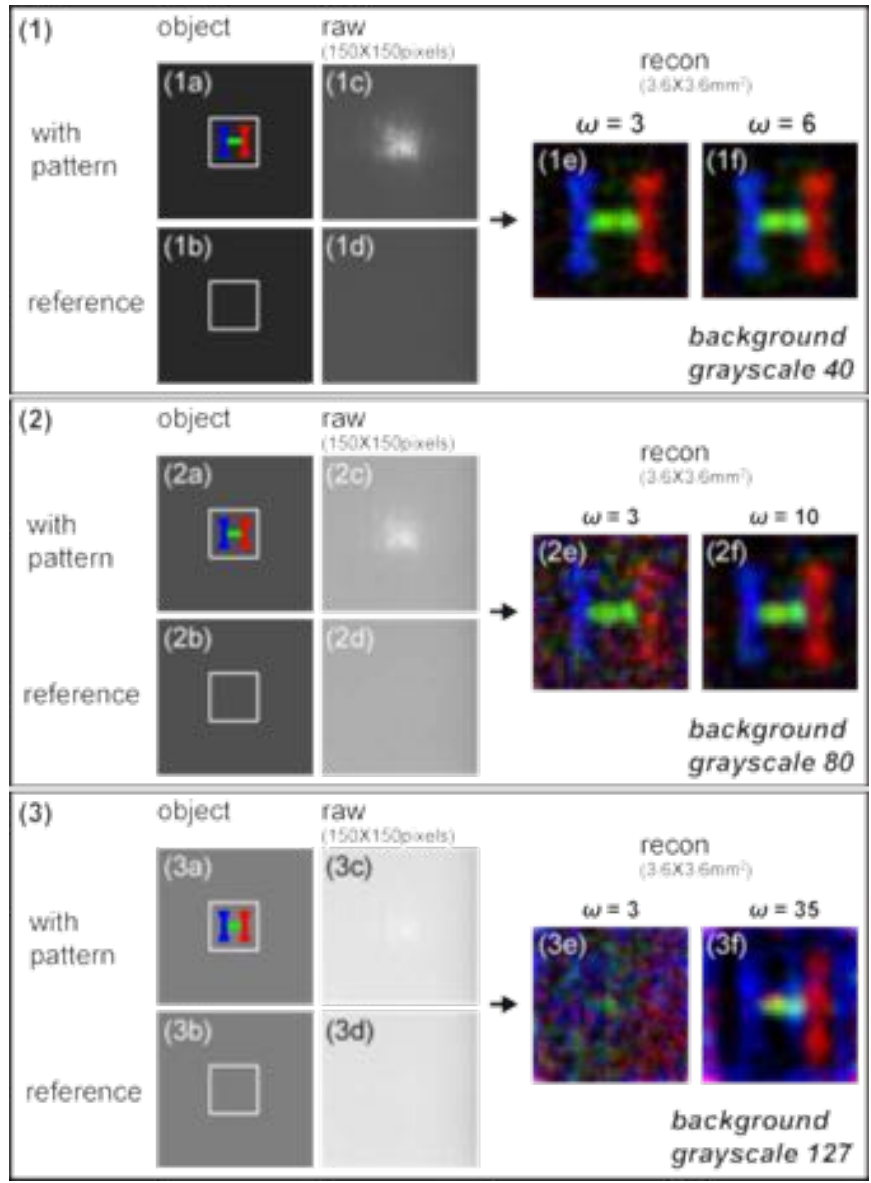

**Figure S19 | Experimental results on object pattern (a three-color letter 'H') with non-black background. The background has grayscale levels of (1) 40, (2) 80 and (3) 127. (a) The original object to be displayed on iPhone screen. (b) The original reference to be displayed on iPhone screen. (c) The raw monochrome image of object (150 by 150 sensor pixels). (d) The raw monochrome image of reference (150 by 150 sensor pixels). (e) Reconstruction images using the regularization parameters ($\omega$=3) the same with all the other results in this work. (f) Reconstruction images using the regularization parameters optimized for different background grayscale levels: (1f) $\omega$=6 for 40; (2f) $\omega$=10 for 80; (3f) $\omega$=35 for 127.**

## 10.    Visible-IR Imaging

Figure S20 gives some calibration PSFs at five exemplary object positions for *λ*=850nm. The PSFs are the same with before for the other 24 wavelengths from 430nm to 706nm with 12nm spacing. Figure S21 gives the raw monochrome image and the reconstructed full multi-spectral data. Here only one single IR laser beam spot projects upon the iPhone screen. However, it is more interesting to introduce some patterns in the IR side. Thus we designed a simplest experiment by cutting the IR laser spot into half by a lab blade at 45°, which is schematically sketched in Fig. S22(a). Figure S22(b) gives both the raw image and the reconstructed multi-spectral images. Here, the relative displacement between the iPhone and the IR beam is shifted. The same green letter 'T' is used as the object pattner.

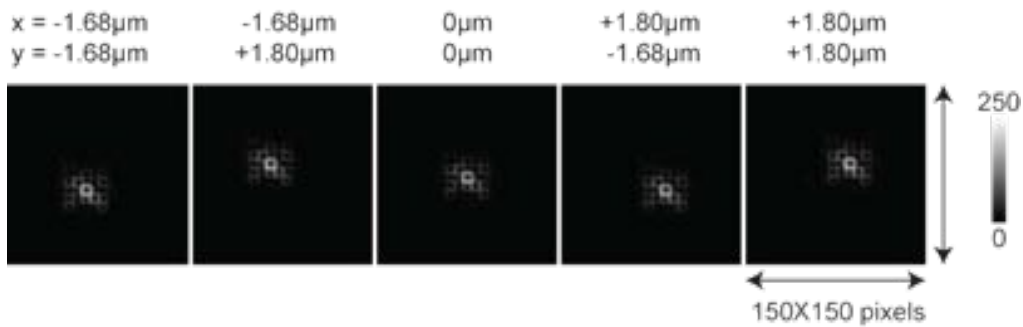

**Figure S20 | Experimentally calibrated PSFs at five object positions for *λ*=850nm. Again, they are cropped to 150 by 150 pixels from the original sensor images.**

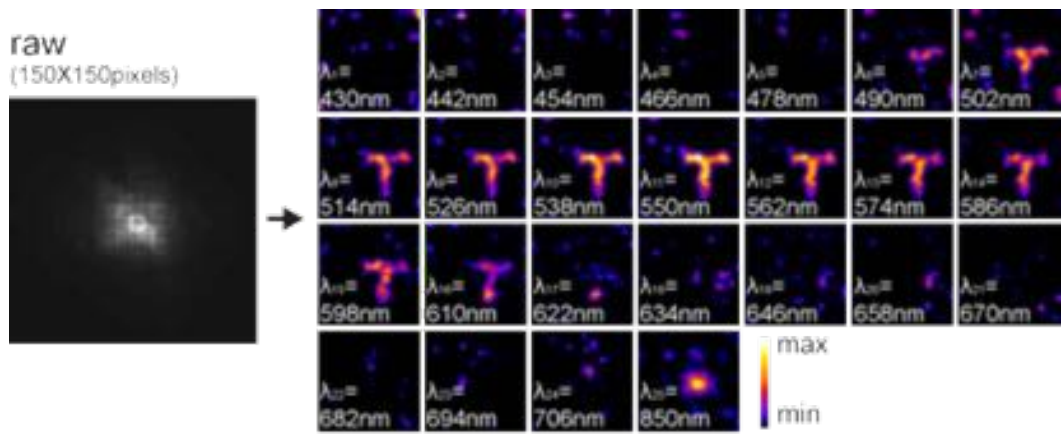

**Figure S21 | Raw monochrome sensor image for visible-IR imaging and the reconstructed multi-spectral images at 25 wavelengths.**

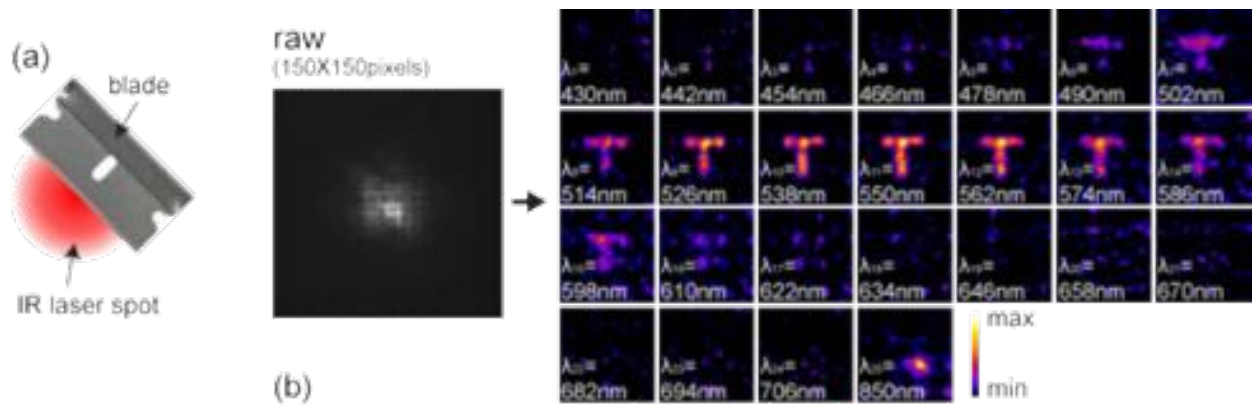

**Figure S22 | (a) Schematic of the IR laser spot cut by a lab blade at 45° to create a simplest pattern to project on the iPhone screen. (b) Raw monochrome sensor image for visible-IR imaging and the reconstructed multi-spectral images at 25 wavelengths.**

## 11.    Spectral Cross-talk

To study the spectral cross-talk of the multi-spectral imager, we reconstructed the objects of single pinhole illuminated by a set of wavelengths from 422nm to 726nm with 4nm spacing, selected by the VARIA tunable bandpass filter (see Sections 2.1 and 2.3 in this Supplementary Materials). The bandwidth is still 12nm. The multi-spectral images are reconstructed as usual and then the spectra at the pinhole center are used to estimate the spectral cross-talk between channels, as plotted in Fig. S23. As can be seen, the cross-talk between channels is trivial except for the longer wavelengths (>700nm), where the sensitivity of silicon-based sensor chip is weak. The average cross-talk between neighbor channels is about -5dB and the average cross-talk between channels that are not neighbors is far less than -10dB.

Figure S23 are the results for visible light imaging, while Fig. S24 contains the cross-talk plots for visible-IR imaging. Here the last wavelength is replaced by 850nm. The reconstructed IR of 850nm is marked by red dot. As expected, the introduction of IR brings negligible cross-talk between the IR wavelengths and all the other bands.

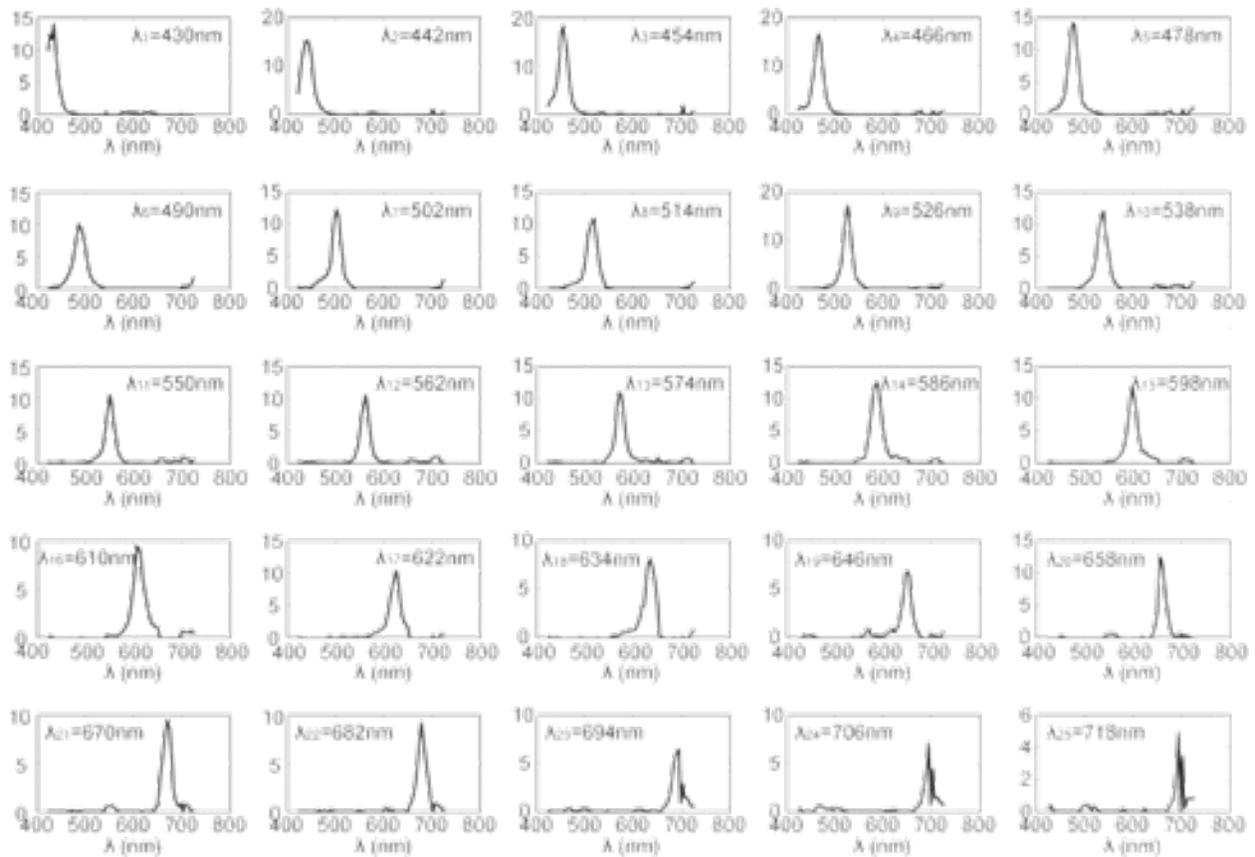

**Figure S23 | Cross-talk plots for visible light imaging. All 25 calibrated wavelengths are tested.**

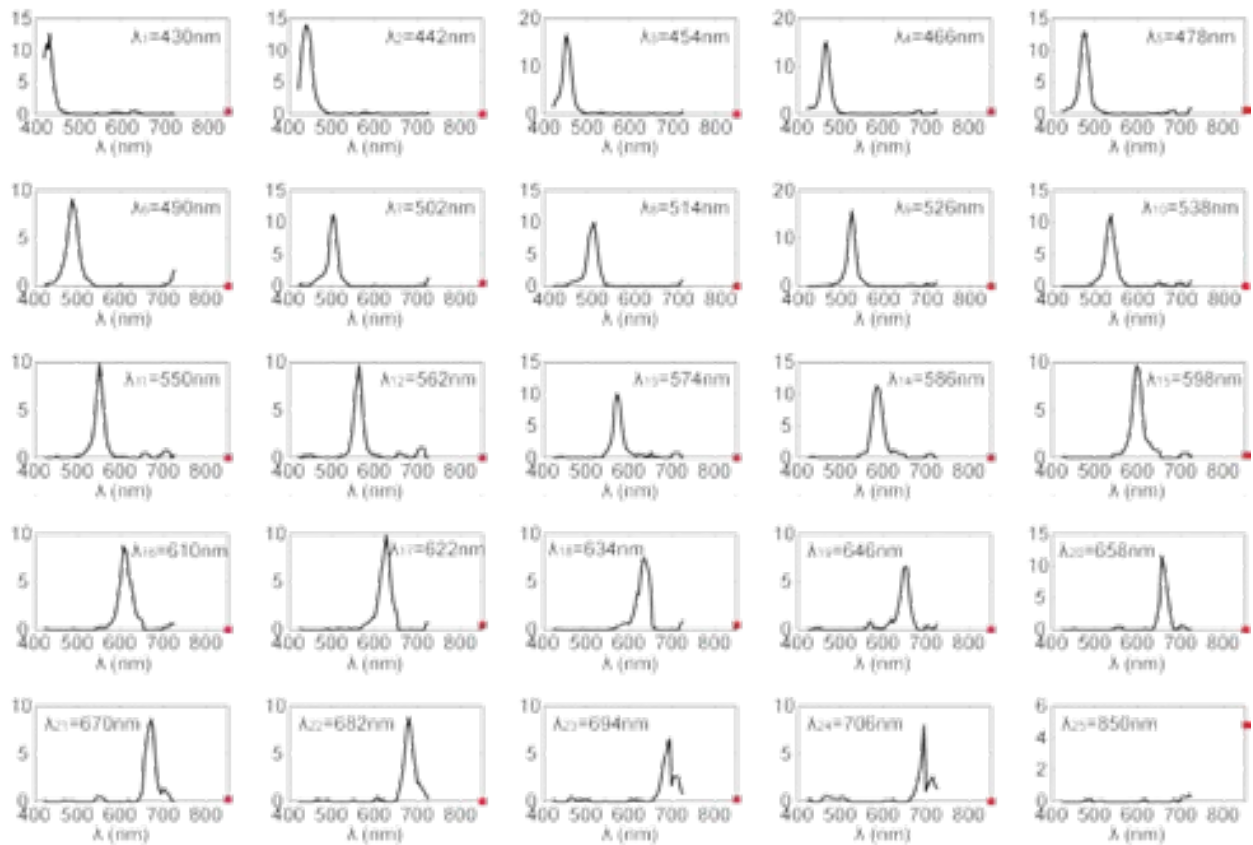

**Figure S24 | Cross-talk plots for visible-IR imaging. All 25 calibrated wavelengths are tested. The response of 850nm IR light is marked by red dots.**

# 12.    Trading off spatial and spectral resolution computationally without changing hardware

In this section, we show that we can use the same DFA and sensor to generate images of size 50 X 50 sensor pixels X 9 wavelength bands by appropriately changing the calibration. In this case, the calibration was conducted for 50 steps in X and Y with a step size of 0.12mm and 9 wavelength bands (430nm to 670nm with bandwidth of 30nm each). So the total field of view is 6mm X 6mm. The resulting multispectral images are summarized in Figs. S25 and S26 below.

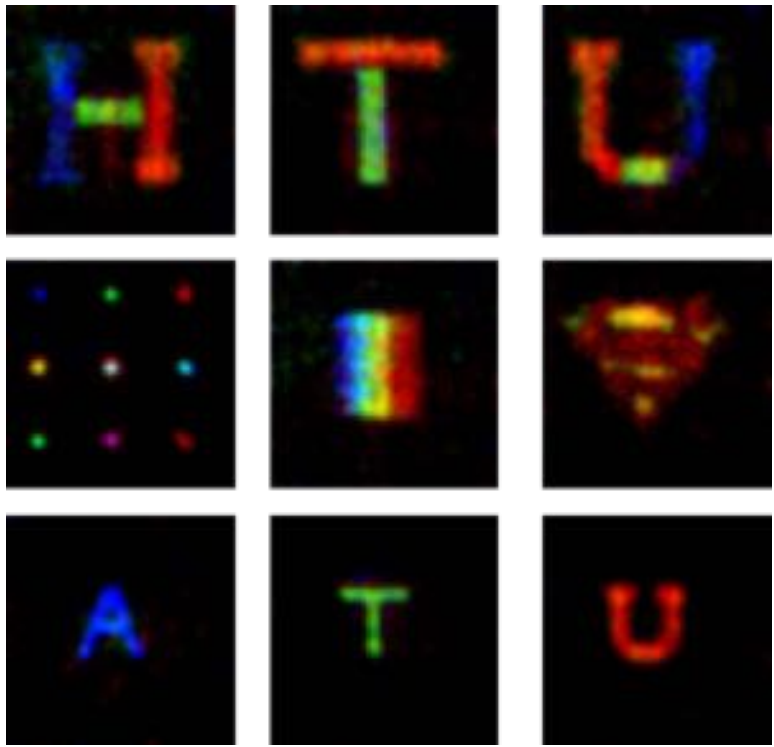

**Figure S25 | Reconstructed multi-spectral images (shown as RGB) with 50 X 50 pixels X 9 wavelength bands. The camera hardware is unchanged compared to the 30 X 30 pixels X 25 bands shown in the main text.**

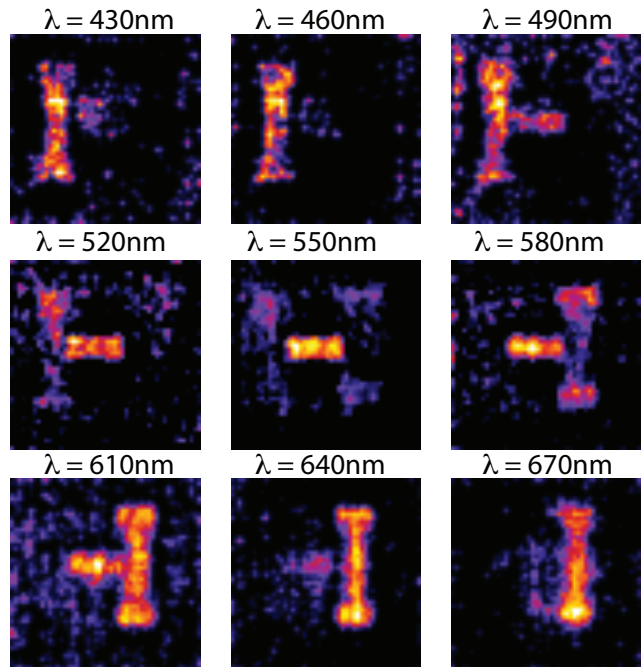

(a)

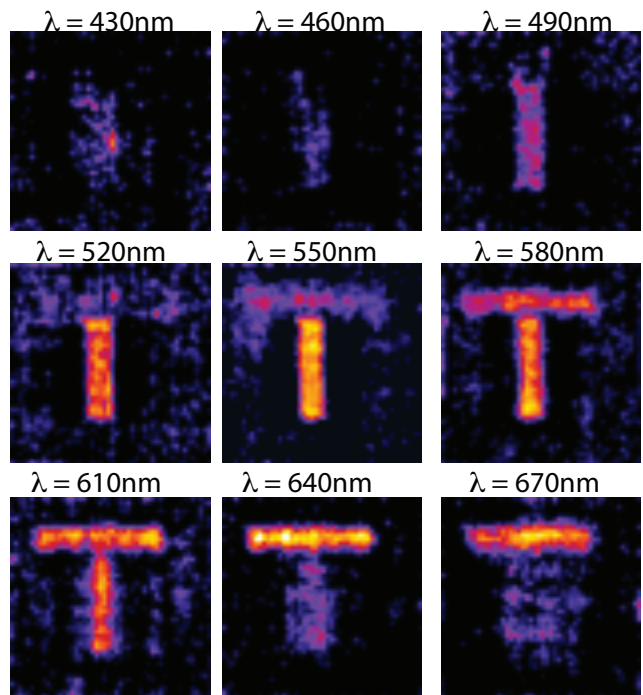

(b)

**Figure S26 | Reconstructed multi-spectral images of (a) 3-color H (RGB image in Fig. S25 a) and 2-color T (RGB image in Fig. S25 b) with 50 X 50 pixels X 9 wavelength bands. The camera hardware is unchanged compared to the 30 X 30 pixels X 25 bands shown in the main text.**